\documentclass[pre,twocolumn,showpacs]{revtex4-1}

\usepackage{amssymb,amsmath}

\usepackage{color}
\newcommand{\vicente}[1]{{ #1}}
\usepackage{graphicx}

\begin{document}

\title{Transport properties for driven granular fluids in situations close to homogeneous steady states}
\author{Vicente Garz\'o}
\email{vicenteg@unex.es} \homepage{http://www.unex.es/eweb/fisteor/vicente/}
\author{Mois\'es G. Chamorro}
\author{Francisco Vega Reyes}
\email{fvega@unex.es} \homepage{http://www.unex.es/eweb/fisteor/fran/}
\affiliation{Departamento de F\'{\i}sica, Universidad de Extremadura, E-06071 Badajoz, Spain}

\begin{abstract}

The transport coefficients of a granular fluid driven by a stochastic bath with friction are obtained by solving the inelastic Enskog kinetic equation from the Chapman-Enskog method. The heat and momentum fluxes as well as the cooling rate are determined to first order in the deviations of the hydrodynamic field gradients from their values in the homogeneous steady state. Since the collisional cooling cannot be compensated locally for the heat produced by the external driving force, the reference distribution $f^{(0)}$ (zeroth-order approximation) depends on time through its dependence on temperature. This fact gives rise to conceptual and practical difficulties not present in the undriven case. On the other hand, to simplify the analysis and given that we are interested in computing transport in the first order of deviations from the reference state, the steady-state conditions are considered to get explicit forms for the transport coefficients and the cooling rate. A comparison with recent Langevin dynamics simulations for driven granular fluids shows an excellent agreement for the kinematic viscosity although some discrepancies are observed for the longitudinal viscosity and the thermal diffusivity at large densities. Finally, a linear stability analysis of the hydrodynamic equations with respect to the homogeneous steady state is performed. As expected, no instabilities are found thanks to the presence of the external bath.

\end{abstract}

\draft 

\date{\today}
\maketitle

\section{Introduction}
\label{sec1}

It is well established that when a granular material is externally excited (rapid flow conditions) the motion of grains resembles the random motion of atoms or molecules in an ordinary gas. In these conditions, kinetic theory together with numerical simulations are the best tools to describe the behavior of granular flows. However, in contrast to ordinary fluids, the collisions between grains are inelastic and so, one has to feed energy into the system to achieve a \emph{steady} non-equilibrium state. This can done either by driving through the boundaries, for example, shearing the system or vibrating its walls \cite{vibration}, or alternatively by bulk driving, as in air-fluidized beds \cite{AD06,SGS05}. On the other hand, this way of supplying energy causes in most of the cases strong spatial gradients in the system. To avoid the difficulties associated with inhomogeneous states, it is quite usual in computer simulations to homogeneously heat the system by the action of an external driving force \cite{thermostat,NETP99,VAZ11,GSVP11}. Borrowing a terminology often used in nonequilibrium molecular dynamics of ordinary fluids \cite{EM90}, this type of external forces are usually called ``thermostats''. Nevertheless, in spite of its practical importance, the understanding of the effect of the external driving force on the dynamical properties of the system (such as the transport coefficients) is still not completely understood \cite{D86,GSB90,GS03}. In particular, recent computer simulations \cite{VAZ11,GSVP11} have computed some transport coefficients by measuring the static and dynamical structure factors for shear and longitudinal modes in a driven granular fluid. Given that the expressions for the transport coefficients were not known in this driven problem, the simulation data were compared with their corresponding elastic forms. Thus, it would be desirable
to provide simulators with the appropriate theoretical tools to work when studying problems in granular fluids driven by thermostats.

The aim of this paper is to determine the transport coefficients of a dense driven granular gas of inelastic hard spheres in the framework of the Enskog kinetic equation. As in the undriven case \cite{GD99,L05}, the transport coefficients are obtained by solving the Enskog equation by means of the Chapman-Enskog expansion \cite{CC70} around a certain reference state $f^{(0)}$ (zeroth-order approximation). While in the undriven case the distribution $f^{(0)}$ is chosen to be the local version of the homogeneous cooling state (HCS), there is some flexibility in the choice of $f^{(0)}$ for a driven gas. For simplicity, one possibility is to take a local thermostat such that the distribution $f^{(0)}$ is still stationary at any point of the system. This was the choice assumed in previous works \cite{GM02,G11} to compute the transport coefficients of a heated granular gas. On the other hand, for \emph{general} small deviations from the steady reference state, the zeroth-order distribution $f^{(0)}$ is not in general a stationary distribution since the collisional cooling cannot be compensated \emph{locally} by the heat injected by the driving force. This fact introduces additional difficulties not present in previous studies \cite{GM02,G11}. In this paper, we will adopt this point of view and will consider this kind of thermostat that seems to be closer to the one used in computer simulations.

The underlying motivation of such a study is twofold. On of the one hand, given that most of the computer simulations \cite{thermostat} are performed by driving the fluid by means of a thermostat, it is important to know the effect of thermostat on the transport properties. In this sense, the question arises then as to whether, and if so to what extent, the conclusions
drawn in previous works \cite{GM02,G11} may be altered when the time-dependent distribution $f^{(0)}(\mathbf{r}, \mathbf{v},t)$ is considered in the Chapman-Enskog expansion instead of the steady distribution. On the other hand, it is also of interest to evaluate the transport coefficients in a situation where a direct comparison with computer simulations is more likely to occur. Here, in the same way as the simulations carried out in Ref.\ \cite{GSVP11}, we have assumed that the fluid is driven by a stochastic bath with friction. This allows us to compare directly our theoretical predictions for the kinematic and longitudinal viscosities and the thermal diffusivity with those obtained from Langevin dynamics simulations \cite{GSVP11}.

The derivation of macroscopic equations and associated transport coefficients from kinetic theory is limited in practice to weakly coupled systems (dilute and moderately dense gases, ideal plasmas, anharmonic crystals). In the case of ordinary fluids, the application of formally exact methods from nonequilibrium statistical mechanics has provided exact expressions for the transport coefficients known as Green-Kubo formula \cite{M89}. Green-Kubo expressions are time integrals of equilibrium time correlation functions for the fluxes associated with conserved quantities. These expressions have the advantage of being formally exact and closely related to properties measured in the experiments. The derivation of the corresponding Green-Kubo formula for the transport coefficients of an \emph{undriven} granular fluid has been a subject of great interest in the past few years \cite{GK}. On the other hand, much less is known in the case of \emph{driven} granular fluids. To the best of our knowledge, the only derivation of Green-Kubo formula for Navier-Stokes transport coefficients of a granular dilute gas heated by the stochastic thermostat has been recently carried out by Garc\'{\i}a de Soria \emph{et al.} \cite{Pablo12}. Starting from the Boltzmann equation, these authors obtain explicit expressions for the transport coefficients as a function of the inelasticity and the spatial dimension. As will show below, in the low-density limit, our results agree with those obtained in Ref.\ \cite{Pablo12} showing the equivalence between the Chapman-Enskog  and linear response methods for driven granular gases.

The plan of the paper is as follows. In Sec.\ \ref{sec2}, the Enskog kinetic equation for a granular gas fluidized by an external force is introduced. Then, before considering inhomogeneous situations, Sec.\ \ref{sec3} analyzes the steady homogeneous state. As in the HCS \cite{NE98}, a scaling solution $\varphi_\text{s}$ is proposed at the steady state. However, the new feature is that the dependence of the reduced distribution $\varphi_\text{s}$ on the granular temperature occurs through two dimensionless parameters (dimensionless velocity and the reduced noise strength) instead of a single parameter (dimensionless velocity) as in the HCS. Once the steady homogeneous distribution is well characterized, Sec.\ \ref{sec4} addresses the Chapman-Enskog expansion around the \emph{unsteady} reference distribution $f^{(0)}(\mathbf{r}, \mathbf{v};t)$. The details of the calculations to first order in the spatial gradients are displayed along several Appendices and the explicit expressions of the transport coefficients and the cooling rate are displayed in Sec.\ \ref{sec5}. In reduced form, they are given in terms of the volume fraction, the coefficient of restitution and the parameters of the thermostat. A comparison with recent Langevin dynamics simulations \cite{GSVP11} for hard disks is done in Sec.\ \ref{sec6} showing an excellent agreement in the case of the kinematic viscosity and some discrepancies for the longitudinal viscosity and the thermal diffusivity, specially at large solid fractions. Section \ref{sec7} is devoted to the linear stability analysis around the steady homogeneous state. As in previous studies \cite{NETP99,GSVP11} based on the elastic forms of the Enskog transport coefficients, our results also indicate that the steady homogeneous state is linearly stable. The paper is closed in Sec.\ \ref{sec8} with a discussion of the results derived here.

\section{Enskog kinetic theory for driven systems}
\label{sec2}

We consider a system of inelastic hard spheres in $d$ dimensions with mass $m$ and diameter $\sigma$. Collisions are characterized by a (constant) coefficient of normal restitution $0<\alpha \leq 1$, with $\alpha=1$ in the elastic limit. As said in the Introduction, in order to maintain a stationary fluidized state, the granular fluid is driven by means of an external force or thermostat that acts locally on each particle \cite{WM96}.  This is a quite usual choice in computer simulations \cite{thermostat}. Under these conditions, the equation of motion for a particle $i$ with peculiar velocity ${\bf V}_i$ can be written as
\begin{equation}
\label{0.1}
m\dot{{\bf V}}_i={\bf F}_i^{\text{th}}(t)+{\bf F}_i^{\text{coll}},
\end{equation}
where ${\bf F}_i^{\text{th}}$ is the thermostat force and ${\bf F}_i^{\text{coll}}$ is the force due to inelastic collisions. Here, ${\bf V}_i={\bf v}_i-{\bf U}$ where $\mathbf{v}_i$ is the velocity of the particle $i$ and $\textbf{U}$ is the \emph{local} mean flow velocity. We assume that ${\bf F}_i^{\text{th}}$ is composed by two different terms: (i) a stochastic force where the particles are randomly kicked between collisions \cite{WM96} and (ii) a viscous drag force which mimics the interaction of the particles with and effective viscous ``bath'' at temperature $T_\text{b}$. More explicitly,
${\bf F}_i^{\text{th}}$ is given by \cite{GSVP11}
\begin{equation}
\label{0.2}
\vicente{
{\bf F}_i^{\text{th}}(t)=-\gamma_\text{b}{\bf V}_i(t)+{\bf F}_i^{\text{st}}(t),
}
\end{equation}
where $\gamma_\text{b}$ is a drag coefficient that defines the characteristic interaction time with the external bath, $\tau_\text{b}^{-1}=\gamma_\text{b}/m$. As usual, the stochastic force ${\bf F}_i^{\text{st}}$ is assumed to have the form of a Gaussian white noise \cite{WM96}:
\begin{equation}
\label{0.3}
\langle {\bf F}_i^{\text{st}}(t) \rangle ={\bf 0}, \quad
\langle {\bf F}_i^{\text{st}}(t) {\bf F}_j^{\text{st}}(t') \rangle =m^2 \xi_\text{b}^2 \delta_{ij}\delta(t-t')
\openone,
\end{equation}
where $\openone$ is the $d\times d$ unit matrix and $\xi_\text{b}^2$ represents the strength of the correlation. The forcing term in the Enskog equation associated to ${\bf F}_i^{\text{st}}$ is represented by a Fokker-Planck operator \cite{NE98} of the form $-\frac{1}{2}\xi_\text{b}^2\partial^2/\partial v^2$. One of the advantages of using the model \eqref{0.1} instead of other kind of thermostats is that the temperature of the thermostat $T_\text{b}$ (different from the kinetic temperature of the fluid $T<T_\text{b}$) is always well defined. In particular, this kind of thermostat is able to equilibrate the system when collisions are elastic. Moreover, a similar external driving force to that of Eq.\ \eqref{0.2} has been recently proposed to model the effect of the interstitial fluid on grains in monodisperse gas-solid suspensions \cite{GTSH12}.

Thus, the corresponding Enskog kinetic equation for the one-particle velocity distribution function $f({\bf r},{\bf v},t)$ reads
\begin{equation}
\partial_{t}f+\mathbf{v}\cdot \mathbf{\nabla}f
-\frac{\gamma_\text{b}}{m} \frac{\partial}{\partial
{\bf v}}\cdot {\bf V} f-\frac{1}{2}\xi_\text{b}^2\frac{\partial^2}{\partial
v^2}f=J_{\text{E}}\left[{\bf r}, {\bf v}|f,f\right],
\label{1}
\end{equation}
where
\begin{eqnarray}
\label{2}
J_{\text{E}}\left[{\bf r}, {\bf v}_{1}|f,f\right] &=&\sigma^{d-1}\int d{\bf v}
_{2}\int d\widehat{\boldsymbol{\sigma}}\,\Theta (\widehat{{\boldsymbol {\sigma }}}
\cdot {\bf g}_{12})(\widehat{\boldsymbol {\sigma }}\cdot {\bf g}_{12})  \nonumber
\\ & \times& \left[ \alpha^{-2}\chi({\bf r},{\bf r}-\boldsymbol {\sigma })
f({\bf r}, {\bf v}_1';t)f({\bf r}-\boldsymbol {\sigma}, {\bf v}_2';t)\right.
\nonumber\\
&- & \left.\chi({\bf r},{\bf r}+\boldsymbol {\sigma }) f({\bf r}, {\bf v}_1;t)f({\bf r}
+\boldsymbol {\sigma }, {\bf v}_2;t)\right]
\end{eqnarray}
is the Enskog collision operator. Here, $d$ is the dimensionality of
the system ($d=2$ for disks and $d=3$ for spheres), $\boldsymbol
{\sigma}=\sigma \widehat{\boldsymbol {\sigma}}$, $\widehat{\boldsymbol
{\sigma}}$ being a unit vector,
$\Theta $ is the Heaviside step function, and ${\bf g}_{12}={\bf v}_{1}-{\bf v}_{2}$. The primes on the velocities in Eq.\ \eqref{2} denote the initial values $\{\mathbf{v}_1', \mathbf{v}_2'\}$ that lead to $\{\mathbf{v}_1, \mathbf{v}_2\}$ following a binary collision, ${\bf v}_{1}^{\prime}={\bf v}_{1}-\frac{1}{2}\left( 1+\alpha^{-1}\right)(\widehat{{\boldsymbol {\sigma }}}\cdot {\bf g}_{12})\widehat{{\boldsymbol {\sigma }}}$, $
{\bf v}_{2}^{\prime }={\bf v}_{2}+\frac{1}{2}\left( 1+\alpha^{-1}\right)
(\widehat{{\boldsymbol {\sigma }}}\cdot {\bf g}_{12})\widehat{
\boldsymbol {\sigma}}$. In addition,
$\chi[{\bf r},{\bf r}+\boldsymbol{\sigma}|\{n(t)] $ is the equilibrium pair correlation function at contact as a functional of the nonequilibrium density field $n({\bf r}, t)$ defined by
\begin{equation}
\label{density}
n({\bf r}, t)=\int\; d{\bf v} f({\bf r},{\bf v},t).
\end{equation}

The macroscopic balance equations for the system are obtained when one
multiplies the Enskog equation (\ref{1}) by $\{1, m{\bf v}, m v^2\}$ and integrates over
velocity. After some algebra one gets \cite{GD99,GTSH12}
\begin{equation}
D_{t}n+n\nabla \cdot {\bf U}=0\;, \label{7}
\end{equation}
\begin{equation}
D_{t}{\bf U}=-\rho ^{-1}\nabla \cdot {\sf P}\;,
\label{8}
\end{equation}
\begin{equation}
D_{t}T+\frac{2}{dn} \left( \nabla \cdot {\bf q}+{\sf P}:\nabla {\bf U}\right) =
-\frac{2 T}{m}\gamma_\text{b} +m \xi_\text{b}^2-\zeta \,T.
\label{9}
\end{equation}
In the above equations, $D_{t}=\partial_{t}+{\bf U}\cdot \nabla$ is
the material derivative and $\rho=m n$ is the mass density. The
cooling rate $\zeta$ is proportional to $1-\alpha^2$ and is due to
dissipative collisions. The pressure tensor ${\sf P}({\bf r},t)$ and
the heat flux ${\bf q}({\bf r},t)$ have both {\em kinetic} and {\em collisional transfer} contributions, i.e., ${\sf P}={\sf P}^k+{\sf P}^c$ and ${\bf q}={\bf q}^k+{\bf q}^c$. The kinetic contributions
are given by
\begin{equation}
\label{10}
{\sf P}^k=\int \; d{\bf v} m{\bf V}{\bf V}f({\bf r},{\bf v},t), \quad
{\bf q}^k=\int \; d{\bf v} \frac{m}{2}V^2{\bf V}f({\bf r},{\bf v},t),
\end{equation}
and the collisional transfer contributions are \cite{GD99}
\begin{eqnarray}
{\sf P}^{c}&=&\frac{1+\alpha}{4}m \sigma^{d}
\int d\mathbf{v}_{1}\int d\mathbf{v}_{2}\int
d\widehat{\boldsymbol {\sigma }}\,\Theta (\widehat{\boldsymbol
{\sigma }}\cdot
\mathbf{g}_{12})(\widehat{\boldsymbol {\sigma }}\cdot \mathbf{g}_{12})^{2} \notag \\
&\times &\widehat{\boldsymbol {\sigma }}\widehat{\boldsymbol {\sigma }}  \int_{0}^{1}dx\; f^{(2)}\left[\mathbf{r}-x{\boldsymbol
{\sigma }},\mathbf{r}+(1-x) {\boldsymbol {\sigma }},\mathbf{v}_{1},\mathbf{v}_{2};t\right],
\nonumber\\
\label{11}
\end{eqnarray}
\begin{eqnarray}
&&{\bf q}^{c}=\frac{1+\alpha}{4}m \sigma^{d}
\int d\mathbf{v}_{1}\int d\mathbf{v}_{2}\int
d\widehat{\boldsymbol {\sigma }}\,\Theta (\widehat{\boldsymbol
{\sigma }}\cdot
\mathbf{g}_{12})(\widehat{\boldsymbol {\sigma }}\cdot \mathbf{g}_{12})^{2}
 \notag \\
& & \times({\bf G}_{12}\cdot\widehat{\boldsymbol {\sigma }})
\widehat{\boldsymbol {\sigma}}\int_{0}^{1}dx\; f^{(2)}\left[\mathbf{r}-
x{\boldsymbol{\sigma}},\mathbf{r}+(1-x)
{\boldsymbol {\sigma}},\mathbf{v}_{1},\mathbf{v}_{2};t\right].\nonumber\\
\label{12}
\end{eqnarray}
Here, ${\bf G}_{12}=\frac{1}{2}({\bf V}_1+{\bf V}_2)$ is the velocity of center of mass and
\begin{equation}
\label{13}
f^{(2)}({\bf r}_1, {\bf r}_2, {\bf v}_1, {\bf v}_2, t)\equiv \chi({\bf r}_1, {\bf r}_2|n(t))f({\bf r}_1, {\bf v}_1, t)
f({\bf r}_2, {\bf v}_2, t).
\end{equation}
Finally, the cooling rate is given by
\begin{eqnarray}
\zeta &=&\frac{\left(1-\alpha^{2}\right)}{4dnT} m \sigma^{d-1}\int d\mathbf{v}
_{1}\int d\mathbf{v}_{2}\int d\widehat{\boldsymbol {\sigma }}
\Theta (\widehat{\boldsymbol {\sigma }}\cdot
\mathbf{g}_{12})\nonumber\\
&&\times(\widehat{ \boldsymbol {\sigma }}\cdot
\mathbf{g}_{12})^{3}f^{(2)}(\mathbf{r}, \mathbf{r}+\boldsymbol {\sigma
},\mathbf{v}_{1},\mathbf{v}_{2};t). \label{14}
\end{eqnarray}

The model \eqref{1} can be seen as the dilute version of the Fokker-Planck model studied previously by Hayakawa \cite{H03} when both model parameters $\gamma_\text{b}$ and $\xi_\text{b}^2$ are related by $\xi_\text{b}^2=2 \gamma_\text{b} T_\text{b}/m^2$ \cite{M89}. Thus, for homogeneous situations, the steady distribution without collisions relaxes to the Maxwellian distribution
\begin{equation}
\label{13.1}
f_\text{M}(V)=n\left(\frac{m}{2\pi T_\text{b}}\right)^{d/2}\; e^{-mV^2/2T_\text{b}}
\end{equation}
where $\mathbf{V}=\mathbf{v}-\mathbf{U}$ is the peculiar velocity.

\section{Steady homogeneous states}
\label{sec3}

Before considering inhomogeneous situations, it is quite instructive to analyze first the homogeneous state. In this situation, the density $n(\mathbf{r}, t)\equiv n_\text{s}$ is constant, the granular temperature $T(\mathbf{r}, t)\equiv T(t)$ is spatially uniform and the mean flow vanishes \vicente{(${\bf U}={\bf 0}$)}. As a consequence, the one-particle distribution function $f({\bf v},t)$ verifies the kinetic equation
\begin{equation}
\partial_{t}f-\frac{\gamma_\text{b}}{m}
\frac{\partial}{\partial{\bf v}}\cdot {\bf v}
f-\frac{1}{2}\xi_\text{b}^2\frac{\partial^2}{\partial v^2}f=J_{\text{E}}[f,f], \label{15}
\end{equation}
where
\begin{eqnarray}
\label{16}
J_{\text{E}}\left[f, f\right] &=&\chi \sigma^{d-1}\int
d\mathbf{v}_{2}\int d\widehat{\boldsymbol {\sigma}}\Theta
(\widehat{\boldsymbol {\sigma}}\cdot \mathbf{g}_{12})(\widehat{
\boldsymbol {\sigma }}\cdot \mathbf{g}_{12})\nonumber\\
&\times&\left[ \alpha^{-2}f(v_{1}^{\prime
})f(v_{2}^{\prime})-f(v_{1})f(v_{2})\right].
\end{eqnarray}
Here, $\chi$ is the pair correlation function evaluated at the (homogeneous) density $n_\text{s}$. The collision operator (\ref{16}) can be recognized as the Boltzmann operator for inelastic collisions multiplied by the factor $\chi$. The energy balance equation reads simply
\begin{equation}
\label{18}
\partial_tT=-\frac{2 T}{m}\gamma_\text{b} +m \xi_\text{b}^2-\zeta \,T.
\end{equation}
In the hydrodynamic regime, $f$ qualifies as a {\em normal} solution and so, its time dependence only occurs through the granular temperature $T$:
\begin{equation}
\label{19}
\partial_t f=\frac{\partial
f}{\partial T}\partial_tT= -\left(\frac{2}{m}\gamma_\text{b} -\frac{m}{T} \xi_\text{b}^2+\zeta\right) T
\frac{\partial f}{\partial T}.
\end{equation}
Substitution of Eq.\ (\ref{19}) into Eq.\ (\ref{15}) yields
\begin{eqnarray}
\label{20} &-&\left( \frac{2}{m}\gamma_\text{b}-\frac{m}{T}\xi_\text{b}^2 + \zeta\right)T
\frac{\partial f}{\partial T}-\frac{\gamma_\text{b}}{m} \frac{\partial}{\partial {\bf
v}}\cdot {\bf v} f\nonumber\\
&-&\frac{1}{2}\xi_\text{b}^2\frac{\partial^2}{\partial
v^2}f=J_{\text{E}}[f,f].
\end{eqnarray}

After a transient regime, the gas will achieve a \emph{steady} state characterized by a constant temperature $T_\text{s}$. According to Eq.\ \eqref{18}, the steady granular temperature $T_\text{s}$ is given by the equation
\begin{equation}
\label{20.1}
\zeta_\text{s} T_\text{s}+ \frac{2\gamma_\text{b}}{m} T_\text{s} =m\xi_\text{b}^2,
\end{equation}
where the (steady) cooling rate $\zeta_\text{s}$ is defined by Eq.\ \eqref{14} by using the stationary distribution function $f_\text{s}(\mathbf{v})$. Equation \eqref{20.1} establishes a relation between the model parameters $\gamma_\text{b}$ and $\xi_\text{b}^2$ so that only one of the above parameters is independent in the steady state. Here, we will take the  noise strength $\xi_\text{b}^2$ as the relevant external parameter. By using the relation \eqref{20.1}, in the steady state Eq.\ \eqref{20} becomes
\begin{equation}
\label{21} \frac{1}{2}\zeta_\text{s} \frac{\partial}{\partial {\bf
v}}\cdot {\bf v} f_\text{s}-\frac{m\xi_\text{b}^2}{2T_\text{s}} \frac{\partial}{\partial {\bf
v}}\cdot {\bf v} f_\text{s}-\frac{1}{2}\xi_\text{b}^2\frac{\partial^2}{\partial v^2}f_\text{s}=J_\text{E}[f_\text{s},f_\text{s}].
\end{equation}

Equation \eqref{21} shows clearly that the steady distribution $f_\text{s}$ also depends on the model parameter $\xi_\text{b}^2$ (apart from its dependence on the coefficient of restitution and the granular temperature). Thus, although the explicit form of $f_\text{s}$ is not known so far, dimensionless analysis requires that $f_\text{s}$ has the scaled form \cite{MVG12}
\begin{equation}
\label{22} f_\text{s}({\bf v}, \xi_\text{b}^2)=n_\text{s}v_0^{-d}\varphi_\text{s}\left(\mathbf{c}, \xi_\text{s}^*\right),
\end{equation}
where $\varphi_\text{s}$ is an unknown function of the dimensionless  parameters
\begin{equation}
\label{22.1}
\mathbf{c}\equiv \frac{\mathbf {v}}{v_0}, \quad \xi_\text{s}^*\equiv \frac{m\ell}{T_\text{s}v_0}\xi_\text{b}^2.
\end{equation}
Here, $v_0=\sqrt{2T_\text{s}/m}$ is the thermal speed and $\ell=1/(n_\text{s}\sigma^{d-1})$ is the mean free path for hard spheres. Note that the dependence of the scaled distribution function $\varphi_\text{s}$ on the temperature is encoded through two parameters: the dimensionless velocity $\mathbf{c}$ and the (reduced) noise strength $\xi_\text{s}^*$. This scaling differs from the one assumed in the case of the HCS \cite{NE98} where only the dimensionless velocity $\mathbf{c}$ is required to characterize the distribution $\varphi_\text{s}$.  A similar scaling solution to the form \eqref{22} has been recently proposed \cite{GMT12} for a driven homogeneous granular gas before reaching the stationary regime. In terms of the (reduced) distribution function $\varphi_\text{s}$, Eq.\ \eqref{21} can be finally rewritten as
\begin{equation}
\label{23} \frac{1}{2}\zeta_\text{s}^* \frac{\partial}{\partial {\bf
c}}\cdot {\bf c}\varphi_\text{s}-\frac{1}{2}\xi_\text{s}^*\frac{\partial}{\partial {\bf c}}\cdot {\bf c} \varphi_\text{s}- \frac{1}{4}\xi_\text{s}^* \frac{\partial^2}{\partial c^2}\varphi_\text{s}=
J_\text{E}^*[\varphi_\text{s},\varphi_\text{s}],
\end{equation}
where we have introduced the dimensionless quantities
\begin{equation}
\label{23.1}
\zeta_\text{s}^*\equiv\frac{\ell\zeta_\text{s}}{v_0}, \quad J_{\text{E}}^*\equiv \frac{\ell v_0^{d-1}}{n_\text{s}}J_{\text{E}}.
\end{equation}

Since the cooling rate vanishes for elastic collisions, then the solution of Eq.\ \eqref{23} is the Maxwellian distribution
\begin{equation}
\label{25}
\varphi_\text{s}({\bf c})= \pi^{-d/2}e^{-c^2}.
\end{equation}
However, if the particles collide inelastically ($\alpha <1$), $\zeta^*\neq 0$ and the exact form of $\varphi_s({\bf c})$ is not known. In particular, the deviation of $\varphi_\text{s}(\mathbf{c}, \xi_\text{s}^*)$ from its Maxwellian form \eqref{25} is measured through the kurtosis or fourth-cumulant
\begin{equation}
\label{28} a_{2,\text{s}}=\frac{4}{d(d+2)}\langle c^4\rangle-1,
\end{equation}
where
\begin{equation}
\label{29} \langle c^k\rangle=\int\; d{\bf c}\; c^k \varphi_s(c).
\end{equation}
The steady-state value $a_{2,\text{s}}$ of the kurtosis can be determined by considering the leading Sonine approximation for $\varphi_\text{s}(\mathbf{c}, \xi_\text{s}^*)$ \cite{NE98}:
\begin{equation}
\label{29.1}
\varphi_\text{s}\simeq   \frac{e^{-c^2}}{\pi^{d/2}}\left\{1 +a_{2,\text{s}}\left[ \frac{c^4}{2}-\frac{(d+2)c^2}{2}+\frac{d(d+2)}{8}\right]\right\},
\end{equation}
The approximation \eqref{29.1} is justified because the coefficient $a_{2,\text{s}}$ is expected to be small \cite{NE98}. With the use of the form \eqref{29.1} and neglecting nonlinear terms in $a_{2,\text{s}}$, the dependence of $a_{2,\text{s}}$ on $\alpha$ and $\xi_\text{s}^*$  can be explicitly determined. After some algebra, one gets \cite{MVG12}
\begin{equation}
\label{30}
a_{2,\text{s}}=\frac{16(1-\alpha)(1-2\alpha^2)}{9+24d-\alpha(41-8d)+30(1-\alpha)\alpha^2
+C_\text{d}\frac{\xi_\text{s}^*}{\chi(1+\alpha)}},
\end{equation}
where $C_\text{d}=16d(d+2)\sqrt{2}\Gamma(d/2)/\pi^{(d-1)/2}$. In the absence of friction ($\gamma_\text{b}=0$), the steady-state condition yields $\zeta_\text{s}^*=\xi_\text{s}^*$ and Eq.\ \eqref{30} agrees with the results obtained when the system is only driven by the stochastic thermostat \cite{NE98}. Moreover, when $\xi_\text{s}^*=0$, we also recover the results of the undriven case  \cite{NE98}.

\begin{figure}
\includegraphics[width=0.85 \columnwidth,angle=0]{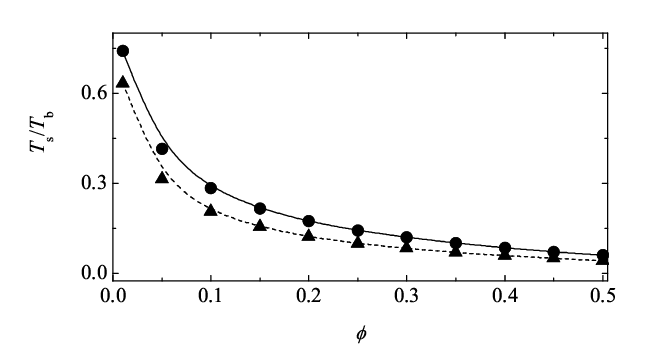}
\caption{Plot of the reduced granular temperature $T_\text{s}/T_\text{b}$ versus the volume fraction
$\phi$ for a two-dimensional ($d=2$) granular fluid and two different values of the coefficient of restitution: $\alpha=0.8$ (solid line) and $\alpha=0.6$ (dashed line). The symbols are the Monte Carlo simulation results (circles for $\alpha=0.8$ and triangles for $\alpha=0.6$).     \label{fig1}}
\end{figure}
\begin{figure}
\includegraphics[width=0.75 \columnwidth,angle=0]{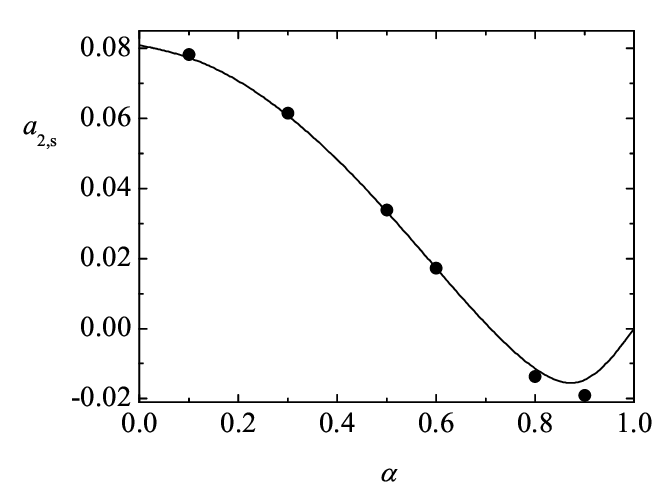}
\caption{Plot of the steady fourth cumulant $a_{2,\text{s}}$ versus the coefficient of restitution $\alpha$ for a two-dimensional ($d=2$) granular fluid with $\phi=0.25$. The line is the theoretical result and the symbols are the Monte Carlo simulation results.     \label{fig2}}
\end{figure}

Once the kurtosis is known, the cooling rate $\zeta_\text{s}$ can be written in terms of $a_{2,\text{s}}$ as
\begin{equation}
\label{31} \zeta_\text{s}=\frac{2}{d}\frac{\pi^{\left( d-1\right) /2}}
{\Gamma \left( \frac{d}{2}\right)}(1-\alpha^2)\chi \left(1+\frac{3}{16}a_{2,\text{s}}\right)
n_\text{s}\sigma^{d-1}\sqrt{\frac{T_\text{s}}{m}},
\end{equation}
where the steady granular temperature $T_\text{s}$ obeys the equation
\begin{equation}
\label{32}
T_\text{s}=\frac{m^2\xi_\text{b}^2}{2\gamma_\text{b}}
-\frac{2^{d-1}}{\sigma}\sqrt{\frac{m}{\pi}}\frac{\chi\phi}
{\gamma_\text{b}}(1-\alpha^2)
\left(1+\frac{3}{16}a_{2,\text{s}}\right)T_\text{s}^{3/2}.
\end{equation}
Here,
\begin{equation}
\label{33}
\phi=
\frac{\pi^{d/2}}{2^{d-1}d\Gamma \left(\frac{d}{2}\right)}n_\text{s}\sigma^d
\end{equation}
is the solid volume fraction. Equation \eqref{32} gives the granular temperature $T_\text{s}$ in the non-equilibrium stationary state.

Figure \ref{fig1} shows the (reduced) steady temperature $T_\text{s}/T_\text{b}$ versus the volume fraction $\phi$ for two different values of the coefficient of restitution $\alpha$. The theoretical results obtained from Eq.\ \eqref{32} for hard disks ($d=2$) are compared with those obtained by numerically solving the Enskog-Boltzmann equation from the direct simulation Monte Carlo (DSMC) method \cite{B94}. As in Ref.\ \cite{GSVP11}, the fixed parameters of the simulations are $m=1$, $\sigma=0.01$, $\gamma_\text{b}=1$, $\xi_\text{b}^2=2$, and $T_\text{b}=1$. For a two-dimensional system, we have chosen the following form for $\chi(\phi)$ \cite{JM87}:
\begin{equation}
\label{33.1}
\chi(\phi)=\frac{1-\frac{7}{16}\phi}{(1-\phi)^2}.
\end{equation}
We observe an excellent agreement between theory and simulation in the complete range of values of $\phi$ considered. As expected, at a given value of the solid fraction, the steady granular temperature $T_\text{s}$ decreases as the gas becomes more inelastic. The dependence of the kurtosis $a_{2,\text{s}}$ on $\alpha$ is shown in Fig.\ \ref{fig2} for $\phi=0.25$, $d=2$, and the same simulation parameters as in Fig.\ \ref{fig1}. It is quite apparent that simulation data compare very well with the theoretical result \eqref{30}, even for extreme values of dissipation. This good agreement suggests that the scaled distribution $\varphi_\text{s} (\mathbf{c},\xi_\text{s}^*)$ can be well represented by the leading Sonine approximation \eqref{29.1} in the region of thermal velocities. In addition, the values of $a_{2,\text{s}}$ in the driven case are generally smaller than in the undriven case \cite{NE98,MS00}.

A similar model (inelastic Enskog equation with a stochastic bath with friction) has been used in Ref.\ \cite{MTC07} to study the nonequilibrium statistical properties of a one dimensional hard rod inelastic fluid. The authors consider a self-consistent density functional approach and compare their theoretical predictions with Brownian dynamics simulations for the hydrodynamic profiles (density, temperature and pressure) and the kurtosis of the velocity distribution function. The agreement found in Ref.\ \cite{MTC07} between theory and simulation is similar to the one observed here in Figs. \ref{fig1} and \ref{fig2} showing the reliability of both approaches (kinetic theory and dynamics density functional theory) to reproduce the steady temperature and the fourth cumulant. However, no attention is devoted in Ref.\ \cite{MTC07} to transport coefficients so that a direct comparison with the latter cannot be carried out in the next Sections.

\section{Small spatial perturbations around the homogeneous steady state}
\label{sec4}

The homogeneous steady state described in the previous Section can be perturbed by small spatial gradients. The response of the system to these perturbations gives rise to nonzero contributions to the heat and momentum fluxes, which are characterized by transport coefficients. The main goal of this paper is to determine the transport coefficients of the \emph{driven} granular fluid. In order to obtain them, we consider states that deviate from steady homogeneous states by \emph{small} spatial gradients. In this case, the Enskog kinetic equation \eqref{1} is solved by means of the Chapman-Enskog method \cite{CC70} conveniently adapted to dissipative dynamics. The Chapman-Enskog method assumes the existence of a normal solution such that all space and time dependence of the distribution function occurs through the hydrodynamic fields
\begin{equation}
f({\bf r},{\bf v},t)=f\left[{\bf v}|n ({\bf r}, t),
T({\bf r}, t), {\bf U}({\bf r}, t) \right] \;.
\label{3.1}
\end{equation}
The notation on the right hand side indicates a functional dependence on the density, temperature and flow velocity. For small spatial variations (i.e., low Knudsen numbers), this functional dependence can be made local in space through an expansion in the gradients of the hydrodynamic fields. To generate it, $f$ is
written as a series expansion in a formal parameter $\epsilon$ measuring the non-uniformity of the system,
\begin{equation}
f=f^{(0)}+\epsilon \,f^{(1)}+\epsilon^2
\,f^{(2)}+\cdots \;, \label{3.2}
\end{equation}
where each factor of $\epsilon$ means an implicit gradient of a
hydrodynamic field. The uniformity parameter $\epsilon$ is related to the Knudsen number defined by the length scale for variation of the hydrodynamic fields. Note that while the strength of the gradients can be controlled by the initial or the boundary conditions in the case of elastic collisions, the problem is more complicated for granular fluids since in some cases (e.g., steady states such as the simple shear flow \cite{G03,SGD04}) there is an intrinsic relation between dissipation and some hydrodynamic gradient. Here, however we consider situations where the spatial gradients are sufficiently small (low Knudsen number). Moreover, in ordering the different level of approximations in the kinetic equation, one has to characterize the magnitude of the external (thermostat) forces relative to the gradients as well. As usual, it is assumed that the external forces (drag and stochastic forces) do not induce any flux in the system and only modify the form of the transport coefficients. As a consequence, $\gamma_\text{b}$ and $\xi_\text{b}^2$ are taken to be of zeroth order in gradients.

According to the expansion (\ref{3.2}) for the distribution function, the Enskog collision operator and time derivative are also expanded in powers of $\epsilon$:
\begin{equation}
\label{3.3}
J_{\text{E}}=J_{\text{E}}^{(0)}+\epsilon J_{\text{E}}^{(1)}+\cdots, \quad \partial_t=\partial_t^{(0)}+\epsilon
\partial_t^{(1)}+\cdots .
\end{equation}
The coefficients in the time derivative expansion are identified by a representation of the fluxes and the cooling rate in the macroscopic balance equations as a similar series through their definitions as functionals of $f$. This is the usual CE method \cite{CC70,GS03} for solving kinetic equations. The expansions \eqref{3.3} lead to similar expansions for the heat and momentum fluxes when substituted into Eqs.\ \eqref{10}--\eqref{12},
\begin{equation}
\label{3.3.1}
P_{ij}=P_{ij}^{(0)}+\epsilon P_{ij}^{(1)}+\cdots, \quad  {\bf q}={\bf q}^{(0)}+\epsilon {\bf q}^{(1)}+\cdots.
\end{equation}
In this paper, we shall restrict our calculations to the first order in the uniformity parameter $\epsilon$.

\subsection{Zeroth-order approximation}

To zeroth order in the expansion, the distribution $f^{(0)}$ obeys the kinetic equation
\begin{equation}
\partial_{t}^{(0)}f^{(0)}-\frac{\gamma_\text{b}}{m}
\frac{\partial}{\partial{\bf v}}\cdot {\bf V}
f^{(0)}-\frac{1}{2}\xi_\text{b}^2\frac{\partial^2}{\partial v^2}f^{(0)}=J_{\text{E}}^{(0)}[f^{(0)},f^{(0)}], \label{3.4}
\end{equation}
where $J_{\text{E}}^{(0)}[f^{(0)},f^{(0)}]$ is given by Eq.\ \eqref{16} with the replacements $n_\text{s}\to n({\bf r},t)$ and $f_\text{s}\to f^{(0)}({\bf r}, {\bf v},t)$. The conservation laws at this order give
\begin{equation}
\label{3.4.1}
\partial_t^{(0)}n=0, \quad \partial_t^{(0)}{\bf U}=\mathbf{0},
\end{equation}
\begin{equation}
\label{3.5}
\partial_t^{(0)}T=-\frac{2T}{m}\gamma_\text{b} +m\xi_\text{b}^2-
\zeta^{(0)}T,
\end{equation}
where $\zeta^{(0)}$ is determined by Eq.\ (\ref{14}) to zeroth order (namely, it is given by Eq.\ \eqref{31} in the first Sonine approximation). The time derivative $\partial_{t}^{(0)}f^{(0)}$ can be more explicitly written as
\begin{eqnarray}
\label{3.5.1}
\partial_t^{(0)} f^{(0)}&=&\frac{\partial
f^{(0)}}{\partial n}\partial_t^{(0)} n+\frac{\partial
f^{(0)}}{\partial U_i}\partial_t^{(0)} U_i+\frac{\partial
f^{(0)}}{\partial T}\partial_t^{(0)} T\nonumber\\
&=&-\left(\frac{2}{m}\gamma_\text{b} -\frac{m}{T} \xi_\text{b}^2+\zeta^{(0)}\right) T\frac{\partial f^{(0)}}{\partial T}.\nonumber\\
\end{eqnarray}
With this result, Eq.\ \eqref{3.4} becomes
\begin{eqnarray}
&-&\left(\frac{2}{m}\gamma_\text{b} -\frac{m}{T} \xi_\text{b}^2+\zeta^{(0)}\right) T\frac{\partial f^{(0)}}{\partial T}-\frac{\gamma_\text{b}}{m}
\frac{\partial}{\partial{\bf v}}\cdot {\bf V}
f^{(0)}\nonumber\\
&-&\frac{1}{2}\xi_\text{b}^2\frac{\partial^2}{\partial v^2}f^{(0)}=J_{\text{E}}^{(0)}[f^{(0)},f^{(0)}]. \label{3.5.2}
\end{eqnarray}

It is important to remark that for given values of $\gamma_\text{b}$, $\xi_\text{b}^2$ and $\alpha$,
the steady-state condition \eqref{20.1} establishes a mapping between the density and temperature so that every density corresponds to one and only one temperature. Since the density $n(\mathbf{r},t)$ and temperature
$T(\mathbf{r},t)$ are specified separately in the \emph{local} reference state $f^{(0)}$, the collisional
cooling is only partially compensated for the heat injected in the system by the external driving force
and so, $\partial_t^{(0)}T\neq 0$. Consequently, the zeroth-order distribution function $f^{(0)}$ depends in general
on time through its dependence on the temperature.
On the other hand, for simplicity, one could impose the steady-state condition \eqref{20.1} at any point of the system and so, $\partial_t^{(0)}T= 0$. This was the choice used in previous theoretical works \cite{GM02,G11} in the case of the stochastic thermostat ($\gamma_\text{b}=0$) where the relation $m\xi_\text{b}^2=\zeta^{(0)} T$ was assumed to apply also in the inhomogeneous state. As we will see below, while the expressions of the shear and bulk viscosities are the same in both choices ($\partial_t^{(0)}T= 0$ and $\partial_t^{(0)}T\neq 0$), the transport coefficients of the heat flux are different. The former choice of thermostat ($\partial_t^{(0)}T\neq 0$) will be referred here to as the choice A while the latter ($\partial_t^{(0)}T= 0$) will be referred as to the choice B. Note that the choice A has the advantage of a simpler implementation in computer simulations. However, at the level of kinetic theory, the fact that $\partial_t^{(0)}T\neq 0$ gives rise to conceptual and practical difficulties not present in the previous analysis \cite{GM02,G11} carried out by using the choice B. The above difficulties are also present in a recent Chapman-Enskog-like method proposed to analyze rheological properties around the steady shear flow state \cite{L06,G06}.

Although for granular gases the drag parameter $\gamma_\text{b}$ and the white noise parameter $\xi_\text{b}^2$ can be considered in general as independent parameters, to make contact here with previous results obtained for dilute gases \cite{Pablo12,H03} we will assume that both parameters are related by
\begin{equation}
\label{3.6}
\gamma_\text{b}=\beta \frac{m^2\xi_\text{b}^2}{T_\text{b}},
\end{equation}
where $\beta$ is an \emph{arbitrary} constant. Thus, when $\beta=0$ our thermostat reduces to the usual stochastic thermostat \cite{Pablo12} while the choice $\beta=\frac{1}{2}$ reduces to the conventional Fokker-Planck model for ordinary gases \cite{H03}. \vicente{According to Eq.\ \eqref{3.6}, the reduced parameter $\gamma^*\equiv \gamma_\text{b}\ell/mv_0$ can be expressed in terms of $\xi^*$ as
\begin{equation}
\label{3.6.1}
\gamma^*=\theta \xi^{*1/3}, \quad \theta\equiv \beta \left(\frac{m\xi_\text{b}^2}{n\sigma^{d-1}T_\text{b}\sqrt{2T_\text{b}/m}}\right)^{2/3}.
\end{equation}
Upon writing \eqref{3.6.1}, use has been made of the identity $\beta T^*=\theta \xi^{*-2/3}$.}

In the \emph{unsteady} state, the zeroth-order distribution function $f^{(0)}$ obeys Eq.\ \eqref{20}. Dimensional analysis requires that $f^{(0)}$ is also given by the scaled form \eqref{22} (once one uses the relation \eqref{3.6}), namely
\vicente{
\begin{equation}
\label{3.7bis} f^{(0)}({\bf r}, {\bf v}, t)=n({\bf r},t) v_0({\bf r},t)^{-d}
\varphi\left(\mathbf{c}, \xi^*, \theta\right),
\end{equation}}
where now $\mathbf{c}\equiv \mathbf{V}/v_0$, $\mathbf{V}={\bf v}-{\bf U}$ being the peculiar velocity. Here, the thermal velocity $v_0$ and the reduced model parameter $\xi^*$ are defined as in Sec.\ \ref{sec3} with the replacement $T_\text{s}\to T(\mathbf{r},t)$. As in the steady state, the temperature dependence of $f^{(0)}$ is not only through $v_0$ and $\mathbf{c}$ but also through $\xi^*$ (see Eq.\ \eqref{22}). Thus,
\begin{equation}
\label{3.8bis} T\frac{\partial f^{(0)}}{\partial T}=-\frac{1}{2}\frac{\partial}{\partial
{\bf V}}\cdot {\bf V} f^{(0)}-\frac{3}{2}\xi^*\frac{\partial f^{(0)}}{\partial
\xi^*},
\end{equation}
and in dimensionless form Eq.\ \eqref{3.4}  can be written as
\begin{eqnarray}
\label{3.7} & & \frac{3}{2}\left[(2\beta T^*-1)\xi^*+\zeta_0^*\right] \xi^*\frac{\partial \varphi}
{\partial \xi^*}+
\frac{1}{2}\left(\zeta_0^*-\xi^*\right) \frac{\partial}{\partial {\bf
c}}\cdot {\bf c} \varphi \nonumber\\
&-& \frac{1}{4}\xi^* \frac{\partial^2}{\partial c^2}\varphi=
J_{\text{E}}^*[\varphi,\varphi],
\end{eqnarray}
where
\begin{equation}
\label{zeta0}
T^*\equiv \frac{T}{T_\text{b}}, \quad \zeta_0^*\equiv \frac{\zeta^{(0)}}{n\sigma^{d-1}\sqrt{2T/m}}.
\end{equation}
Upon writing Eq.\ \eqref{3.7} use has been made of the relation \eqref{3.6}. Note that the reduced temperature $T^*\propto \xi^{*-2/3}$.

As before, the explicit form of $\varphi$ is not known. An indirect information on the scaled distribution $\varphi$ is given through its fourth-cumulant $a_2(\xi^*)$ which is defined by Eq.\ \eqref{28}. This cumulant can be obtained by multiplying both sides of Eq.\ \eqref{3.7} by $c^4$ and integrating over velocity. The result is
\begin{widetext}
\begin{equation}
\label{3.8}
-\frac{3d(d+2)}{8}\left[(2\beta T^*-1)\xi^*+\zeta_0^*\right]\xi^*\frac{\partial a_2}
{\partial \xi^*}+\frac{d(d+2)}{2}\left[\zeta_0^*(1+a_2)-\xi^*a_2\right]=\mu_4,
\end{equation}
\end{widetext}
where
\begin{equation}
\label{3.9.1} \mu_\ell=-\int\; d{\bf c}\;c^\ell\; J_{\text{E}}^*[\varphi,\varphi].
\end{equation}
In the steady-state, Eq.\ \eqref{20.1} applies and the first term on the left hand side of Eq.\ \eqref{3.8}
vanishes. In this case,
the solution to Eq.\ \eqref{3.8} is given by Eq.\ \eqref{30}. In general, Eq.\ \eqref{3.8} must be solved numerically to get the dependence of $a_2$ on $\xi^*$ (or equivalently, on the reduced temperature $T^*$). An analytical expression of $\partial a_2/\partial \xi^*$ in the steady state has been obtained in the Appendix \ref{appA}. Thus, in what follows $a_2(\xi^*)$ will be considered as a known function of $\xi^*$.

\section{Transport coefficients}
\label{sec5}

The analysis to first order in spatial gradients is similar to the one worked out  in the undriven case \cite{GD99,L05,G12}. Some technical details on the determination of the transport coefficients and the cooling rate are provided in the Appendices \ref{appB} and \ref{appC}. The form of the first-order velocity distribution function $f^{(1)}$ is given by
\begin{eqnarray}
\label{5.1}
f^{(1)}&=&\boldsymbol{\mathcal{A}}\left(
\mathbf{V}\right)\cdot  \nabla \ln
T+\boldsymbol{\mathcal{B}}\left(
\mathbf{V}\right) \cdot \nabla \ln n
\nonumber\\
& &+ \mathcal{C}_{ij}\left( \mathbf{V} \right)\frac{1}{2}\left( \partial _{i}U_{j}+\partial _{j
}U_{i}-\frac{2}{d}\delta _{ij}\nabla \cdot
\mathbf{U} \right)\nonumber\\
& & +\mathcal{D}\left( \mathbf{V} \right) \nabla \cdot
\mathbf{U},
\end{eqnarray}
where the quantities $\boldsymbol{\mathcal{A}}\left(\mathbf{V}\right)$, $\boldsymbol{\mathcal{B}}\left(
\mathbf{V}\right)$, $\mathcal{C}_{ij}\left( \mathbf{V} \right)$ and $\mathcal{D}\left( \mathbf{V} \right)$ are the solutions of the linear integral equations \eqref{b13}--\eqref{b16}, respectively. However, the evaluation of the transport coefficients from the above integral equations requires to know the complete time dependence of the first order contributions to the pressure tensor and the heat flux vector. This is quite an intricate problem. On the other hand, some simplifications occur if attention is restricted to linear deviations from the steady state described in Section \ref{sec2}. In particular, since the kinetic and collisional contributions to the heat and momentum fluxes are already of first order in the deviations from the steady state, one only needs to know the transport coefficients to zeroth order in the deviations. This means we can evaluate the transport coefficients in the steady state conditions, namely, when the condition \eqref{20.1} applies.

In this case, Eqs.\ \eqref{b13}--\eqref{b16} become
\begin{eqnarray}
\label{5.2}
&-&\left[
\frac{m}{T}\xi_\text{b}^2\left(1-\frac{3}{2}\frac{\partial \zeta_0^*}{\partial \xi^*}\right)+\frac{1}{2}\zeta^{(0)}\right]
\boldsymbol{\mathcal{A}}-\frac{\gamma_\text{b}}{m}\frac{\partial}{\partial {\bf v}}\cdot
{\bf V}\boldsymbol{\mathcal{A}}\nonumber\\
&-&\frac{1}{2}\xi_\text{b}^2
\frac{\partial^2}{\partial v^2}\boldsymbol{\mathcal{A}}+{\cal L}\boldsymbol{\mathcal{A}}={\bf A},
\end{eqnarray}
\vicente{
\begin{eqnarray}
\label{5.3}
&-&
\frac{\gamma_\text{b}}{m}\frac{\partial}{\partial {\bf v}}\cdot {\bf V}\boldsymbol{\mathcal{B}}
-\frac{1}{2}\xi_\text{b}^2
\frac{\partial^2}{\partial v^2}\boldsymbol{\mathcal{B}}+{\cal L}\boldsymbol{\mathcal{B}}={\bf B}
\nonumber\\
&+&
\zeta ^{(0)}\left(1+\phi\frac{\partial}{\partial \phi}\ln \chi \right)\boldsymbol{\mathcal{A}}\nonumber\\
&+&\left[\phi \frac{\partial \chi}{\partial \phi}\frac{\partial}{\partial \chi}\left(\frac{\zeta^{(0)}}{\chi}\right)-\xi^*\frac{\partial \zeta^{(0)}}{\partial \xi^*}
-\frac{2}{3}\theta\frac{\partial \zeta^{(0)}}{\partial \theta}\right]
\boldsymbol{\mathcal{A}},
\end{eqnarray}}
\begin{equation}
-\frac{\gamma_\text{b}}{m}\frac{\partial}{\partial {\bf v}}\cdot {\bf V}\mathcal{C}_{ij}
-\frac{1}{2}\xi_\text{b}^2
\frac{\partial^2}{\partial v^2}\mathcal{C}_{ij}+{\cal L}\mathcal{C}_{ij}=C_{ij},  \label{5.4}
\end{equation}
\begin{equation}
-\frac{\gamma_\text{b}}{m}\frac{\partial}{\partial {\bf v}}\cdot {\bf V}\mathcal{D}
-\frac{1}{2}\xi_\text{b}^2
\frac{\partial^2}{\partial v^2}\mathcal{D}+{\cal L}\mathcal{D}=D,  \label{5.5}
\end{equation}
where it is understood that the quantities $\mathbf{A}$, $\mathbf{B}$, $C_{ij}$, and $D$ (defined by Eqs.\ \eqref{b5}--\eqref{b8}, respectively) are evaluated in the steady state. Consequently, \emph{all} the transport coefficients are given in terms of the steady granular temperature $T_\text{s}$.

The forms of the collisional contributions to the momentum and heat fluxes are exactly the same as those previously obtained in the undriven case \cite{GD99,L05} except that $a_{2,\text{s}}$ depends on $\xi_\text{s}^*$ (see Eq.\ \eqref{30}). Thus, we will focus here our attention in the evaluation of the kinetic contributions to the transport coefficients and the cooling rate. Technical details on this calculation are given in the Appendix \ref{appC}.

\subsection{Pressure tensor}

To first order in the spatial gradients, the pressure tensor is given by
\begin{equation}
\label{5.6}
P_{ij}^{(1)}=-\eta\left( \partial_{i}U_{j}+\partial _{j
}U_{i}-\frac{2}{d}\delta _{ij}\nabla \cdot
\mathbf{U} \right) - \lambda \delta_{ij} \nabla \cdot \mathbf{U},
\end{equation}
where $\eta$ is the shear viscosity and $\lambda$ is the bulk viscosity. While the shear viscosity has kinetic and collisional contributions, the bulk viscosity has only a collisional contribution. The bulk viscosity $\lambda$ is given by
\begin{equation}
\label{5.7}
\lambda=\frac{2^{2d+1}}{\pi(d+2)}\phi^2 \chi (1+\alpha)\left(1-\frac{a_{2,\text{s}}}{16} \right)\eta_0,
\end{equation}
where
\begin{equation}
\label{5.8}
\eta_0=\frac{d+2}{8}\frac{\Gamma \left( \frac{d}{2}\right)}{\pi ^{\left( d-1\right) /2}}\sigma^{1-d}\sqrt{mT_\text{s}}
\end{equation}
is the low density value of the shear viscosity in the elastic limit. The shear viscosity $\eta$ can be written as
\begin{eqnarray}
\label{5.9}
\eta&=&\frac{\eta_0\nu_0}{\nu_\eta+\frac{2\beta m}{T_\text{b}}\xi_\text{b}^2}
\left[1-\frac{2^{d-2}}{d+2}(1+\alpha)
(1-3 \alpha)\phi \chi \right]\nonumber\\
& \times& \left[1+\frac{2^{d-1}}{d+2}(1+\alpha)
\phi \chi \right]+\frac{d}{d+2}\lambda,
\end{eqnarray}
where $\nu_0=n_\text{s}T_\text{s}/\eta_0$ and the collision frequency $\nu_\eta$ is \cite{GSM07}
\begin{equation}
\label{5.10}
\nu_\eta=\frac{3\nu_0}{4d}\chi \left(1-\alpha+\frac{2}{3}d\right)(1+\alpha)
\left(1+\frac{7}{16}a_{2,\text{s}}\right).
\end{equation}

\subsection{Heat Flux}

The constitutive form for the heat flux in the Navier-Stokes approximation is
\begin{equation}
\label{5.11}
{\bf q}^{(1)}=-\kappa \nabla T-\mu \nabla n,
\end{equation}
where $\kappa$ is the thermal conductivity and $\mu$ is a new coefficient not present in the elastic case ($\alpha=1$).

The thermal conductivity $\kappa$ is given by
\begin{eqnarray}
\label{5.12}
\kappa&=&\kappa_k\left[1+3\frac{2^{d-2}}{d+2}\phi \chi (1+\alpha)\right]
\nonumber\\
&+&\frac{2^{2d+1}(d-1)}{(d+2)^2\pi}
\phi^2 \chi (1+\alpha)\left(1+\frac{7}{16} a_{2,\text{s}} \right)\kappa_0,
\end{eqnarray}
where
\begin{equation}
\label{5.13}
\kappa_0=\frac{d(d+2)}{2(d-1)}\frac{\eta_0}{m}
\end{equation}
is the thermal conductivity coefficient of an elastic dilute gas. The expression of the kinetic part $\kappa_k$ appearing
in Eq.\ \eqref{5.12} is
\begin{eqnarray}
\label{5.14}
\kappa_k&=&\kappa_0\nu_0\frac{d-1}{d}\left[\nu_\kappa+\frac{1}{2}\frac{m\xi_\text{b}^2}{T_\text{s}}\left(1+3 \zeta_\text{M}
\left(\frac{\partial a_2}{\partial \xi^*}\right)_\text{s}\right)-2\zeta_\text{s}^{(0)}
\right]^{-1}
\nonumber\\
& &\left\{1+2a_{2,s}-\frac{3}{2}\xi_\text{s}^*\left(\frac{\partial a_2}{\partial \xi^*}\right)_\text{s}+3\frac{2^{d-3}}{d+2}\phi \chi(1+\alpha)^2\left[2\alpha-1\right.\right.\nonumber\\
& & \left.\left.+a_{2,\text{s}}(1+\alpha)\vicente{-\frac{3}{8}(1+\alpha)}
\xi_\text{s}^*\left(\frac{\partial a_2}{\partial \xi^*}\right)_\text{s}\right]\right\}.\nonumber\\
\end{eqnarray}
In Eq.\ \eqref{5.14}, $\zeta_\text{s}^{(0)}$ is given by Eq.\ \eqref{31},
\begin{equation}
\label{5.14.1}
\zeta_\text{M}=\frac{3\sqrt{2}}{16d}\frac{\pi^{\left( d-1\right) /2}}
{\Gamma \left( \frac{d}{2}\right)}(1-\alpha^2)\chi,
\end{equation}
and the value of the derivative $(\partial a_2/\partial \xi^*)_\text{s}$ in the steady-state is provided in the Appendix \ref{appA}. Moreover, the collision frequency $\nu_\kappa$ is  \cite{GSM07}
\begin{eqnarray}
\label{5.15}
\nu_\kappa&=&\nu_0\frac{1+\alpha}{d}\chi\left[\frac{d-1}{2}+\frac{3}{16}(d+8)(1-\alpha)\right.\nonumber\\
& & \left.+\frac{296+217d-3(160+11d)\alpha}{256}a_{2,\text{s}}\right].
\end{eqnarray}

The coefficient $\mu$ is
\begin{equation}
\label{5.16}
\mu=\mu_k\left[1+3\frac{2^{d-2}}{d+2}\phi \chi (1+\alpha)\right],
\end{equation}
\vicente{
where its kinetic contribution $\mu_k$ is
\begin{eqnarray}
\label{5.17}
\mu_k&=&\frac{\kappa_0\nu_0T_\text{s}}{n_\text{s}}\left[\nu_\kappa-\frac{3}{2}\left(\zeta_\text{s}^{(0)}
-\frac{m\xi_\text{b}^2}{T_\text{s}}\right)\right]^{-1}\nonumber\\
& & \left\{\frac{\kappa_k}{\kappa_0\nu_0}\left[
\zeta_\text{s}^{(0)}\left(1+\phi\partial_{\phi}
\ln \chi \right)\right.\right.\nonumber\\
& & \left.+\frac{\zeta_\text{M}v_0}{\ell}\left(\frac{\phi}{\chi}\frac{\partial \chi}{\partial \phi}
\frac{\partial a_2}{\partial \chi}-\xi_\text{s}^*\frac{\partial a_2}{\partial \xi^*}
-\frac{2}{3}\theta\frac{\partial a_2}{\partial \theta}\right)\right]\nonumber\\
& & \left.+\frac{d-1}{d}\left(a_{2,\text{s}}+\phi \frac{\partial \chi}{\partial \phi}
\frac{\partial a_2}{\partial \chi}-\xi_\text{s}^*\frac{\partial a_2}{\partial \xi^*}
-\frac{2}{3}\theta\frac{\partial a_2}{\partial \theta}\right)\right.\nonumber\\
& &
+3\frac{2^{d-2}(d-1)}{d(d+2)}\phi \chi
(1+\alpha)\left(1+\frac{1}{2}\phi\partial_\phi\ln
\chi\right)\nonumber\\
& & \left.\times\left[\alpha(\alpha-1)+\frac{a_{2,\text{s}}}{6}(10+2d-3\alpha+3\alpha^2)\right]\right.
\nonumber\\
& &\left.+3\frac{2^{d-4}(d-1)}{d(d+2)}\chi \phi (1+\alpha)^3\right.
\nonumber\\
& & \left.\times\left(\phi \frac{\partial \chi}{\partial \phi}
\frac{\partial a_2}{\partial \chi}-\xi^*\frac{\partial a_2}{\partial \xi^*}
-\frac{2}{3}\theta\frac{\partial a_2}{\partial \theta}\right)\right\}.
\end{eqnarray}}

\subsection{Cooling rate}

The cooling rate $\zeta$ is given by
\begin{equation}
\label{5.18} \zeta=\zeta_\text{s}^{(0)}+\zeta_U \nabla \cdot {\bf U}.
\end{equation}
At first order in spatial gradients, the proportionality constant $\zeta_U$ is a new transport coefficient for granular fluids \cite{GD99,L05}. For a driven gas, $\zeta_U$ can be written as
\begin{equation}
\label{5.19} \zeta_U=\zeta_{10}+\zeta_{11},
\end{equation}
where
\begin{equation}
\label{5.20} \zeta_{10}= -3\frac{2^{d-2}}{d}\chi \phi (1-\alpha^2),
\end{equation}
\begin{eqnarray}
\label{5.21}
\zeta_{11}&=&\frac{9(d+2)\;2^{d-8}}{d^2}\chi(1-\alpha^2)
\left(\nu_\gamma+\frac{2m\xi_\text{b}^2}{T}-2\zeta_\text{s}^{(0)}\right)^{-1}\nonumber\\
&&\left[\frac{\omega \phi \chi}{2(d+2)}-2^{2-d}\frac{d+3}{3}\xi_\text{s}^*\left(\frac{\partial a_2}{\partial \xi^*}\right)_\text{s}\nu_0\right.\nonumber\\
& & \left.
-(1+\alpha)\left(\frac{1}{3}-\alpha\right)\left(2a_{2,\text{s}}-\frac{3}{2}
\xi_\text{s}^*\left(\frac{\partial a_2}{\partial \xi^*}\right)_\text{s}\right)\phi\chi \nu_0\right]
,\nonumber\\
\end{eqnarray}
and the collision frequencies $\omega$ and $\nu_\gamma$ are
\begin{eqnarray}
\label{5.22}
\omega&=&(1+\alpha)\nu_0\left\{
(1-\alpha^2)(5\alpha-1)\right.\nonumber\\
&-& \left.\frac{a_{2,\text{s}}}{6}\left[15\alpha^3-3\alpha^2+3(4d+15)\alpha-
(20d+1)\right]\right\},\nonumber\\
\end{eqnarray}
\begin{eqnarray}
\label{5.23}
\nu_\gamma&=&-\frac{1+\alpha}{192}\chi \nu_0\left[30\alpha^3-30\alpha^2+(105+24 d) \alpha\right.\nonumber\\
& & \left.-56d-73\right].
\end{eqnarray}
In Eq.\ \eqref{5.21}, for the sake of simplicity, the contributions proportional to the derivatives $\partial a_2/\partial \chi$ and $\partial a_2/\partial \theta$ have been neglected.  Note that the first-order contribution $\zeta_U$ to the cooling rate vanishes for elastic gases ($\alpha=1$, arbitrary solid volume fraction $\phi$). However, for dilute inelastic gases ($\phi=0$, arbitrary values of the coefficient of restitution $\alpha$), at variance with the undriven case \cite{BDKS98} there is here a nonzero contribution to $\zeta_{U}$ proportional to $(\partial a_{2}/\partial \xi^*)_\text{s}$ [see Eq.\ \eqref{5.21}]. This result is consistent with those obtained \cite{Pablo12} from the Boltzmann equation.

The expressions for the Navier-Stokes transport coefficients obtained by using the choice B (i.e., when the condition \eqref{3.5} holds locally and so, $\partial_t^{(0)}T=0$) are displayed in the Appendix \ref{appD}. While the expressions of $\eta$ and $\lambda$ are also given by Eqs.\ \eqref{5.7}--\eqref{5.9}, the forms of $\kappa$ and $\mu$ are different to those derived from the choice A.

\subsection{Some special limits}

It is quite apparent that the expressions of the transport coefficients are rather complicated, given the different parameters (inelasticity, density and the model parameter $\xi_\text{b}^2$) involved in them. Thus, in order to show more clearly the dependence of each parameter on transport, it is instructive to consider some simple cases.

In the elastic limit ($\alpha=1$), $T_\text{s}=m^2\xi_\text{b}^2/2\gamma_\text{b}$, $\zeta_\text{s}^{(0)}=a_{2,\text{s}}=0$, $\nu_\eta=\chi \nu_0$, and $\nu_\kappa=(1-d^{-1})\chi \nu_0$. In this case, $\mu=\zeta_U=0$ and the coefficients $\lambda$, $\eta$ and $\kappa$ become, respectively,
\begin{equation}
\label{5.23.1}
\lambda=\frac{2^{2(d+1)}}{\pi(d+2)}\phi^2 \chi \eta_0,
\end{equation}
\begin{equation}
\label{5.23.2}
\eta=\frac{\eta_0}{\chi+\frac{2\beta m}{T_\text{b}\nu_0}\xi_\text{b}^2}
\left(1+\frac{2^{d}}{d+2}\phi \chi \right)^2+\frac{d}{d+2}\lambda,
\end{equation}
\begin{equation}
\label{5.23.3}
\kappa=\kappa_0\frac{\left(1+3\frac{2^{d-1}}{d+2}\phi \chi \right)^2}{\chi+\frac{d}{d-1}\frac{\gamma_\text{b}}{m\nu_0}}
+\frac{2^{2(d+1)}(d-1)}{(d+2)^2\pi}
\phi^2 \chi\kappa_0.
\end{equation}

Note that the expressions \eqref{5.23.2} and \eqref{5.23.3} for $\eta$ and $\kappa$ differ from their corresponding elastic counterparts for undriven gases.

We consider now a low-density granular gas ($\phi=0$). In this limit case, $\lambda=0$ while $\eta$, $\kappa$ and $\mu$ are given, respectively, by
\begin{equation}
\label{5.9.dilute}
\eta=\frac{\eta_0\nu_0}{\nu_\eta+\frac{2\beta m}{T_\text{b}}\xi_\text{b}^2},
\end{equation}
\begin{eqnarray}
\label{5.14.dilute}
\kappa&=&\kappa_0\nu_0\frac{d-1}{d}\left[\nu_\kappa+\frac{1}{2}\frac{m\xi_\text{b}^2}{T_\text{s}}\left(1+3 \zeta_\text{M}
\left(\frac{\partial a_2}{\partial \xi^*}\right)_\text{s}\right)-2\zeta_\text{s}^{(0)}
\right]^{-1}
\nonumber\\
& &\times\left[1+2a_{2,s}-\frac{3}{2}\xi_\text{s}^*\left(\frac{\partial a_2}{\partial \xi^*}\right)_\text{s}
\right],
\end{eqnarray}
\begin{eqnarray}
\label{5.17.dilute}
\mu&=&\frac{\kappa_0\nu_0T_\text{s}}{n_\text{s}}\left[\nu_\kappa-\frac{3}{2}\left(\zeta_\text{s}^{(0)}
-\frac{m\xi_\text{b}^2}{T_\text{s}}\right)\right]^{-1}\nonumber\\
& \times& \left\{\frac{\kappa}{\kappa_0\nu_0}\left[
\zeta_\text{s}^{(0)}-\frac{\zeta_\text{M}v_0}{\ell}\left(\xi_\text{s}^*\frac{\partial a_2}{\partial \xi^*}
+\frac{2}{3}\theta\frac{\partial a_2}{\partial \theta}\right)\right]
\right.\nonumber\\
& &\left. 
+\frac{d-1}{d}\left(a_{2,\text{s}}-\xi_\text{s}^*\frac{\partial a_2}{\partial \xi^*}
-\frac{2}{3}\theta\frac{\partial a_2}{\partial \theta}\right)\right\},
\end{eqnarray}
where $\nu_\eta$ and $\nu_\kappa$ are defined by Eqs.\ \eqref{5.10} and \eqref{5.15}, respectively, with $\chi=1$.  The expressions \eqref{5.9.dilute} and \eqref{5.14.dilute} agree with recent results \cite{Pablo12} derived from the linearized Boltzmann equation for a granular gas heated by the stochastic thermostat ($\beta=0$). In addition, as mentioned before, when $\beta=\frac{1}{2}$ in Eq.\ \eqref{3.6}, our model reduces to the Fokker-Planck model studied previously by Hayakawa \cite{H03} for dilute gases. In this paper, Hayakawa determines the transport coefficients $\eta$, $\kappa$, and $\mu$ by neglecting the dependence of the fourth cumulant $a_2$ on the (reduced) model parameters $\gamma^*$ and $\xi^*$. In particular,  in the steady state, Eqs.\ \eqref{5.9.dilute}--\eqref{5.17.dilute} agree with the results obtained in Ref.\ \cite{H03} when $(\partial a_2/\partial \xi^*)_\text{s}=0$. All the above limit situations confirm the self-consistency of the results derived here for a dense granular fluid.

\section{Comparison with computer simulations}
\label{sec6}

The expressions derived in Sec.\ \ref{sec5} for the transport coefficients and the cooling rate depend on the (steady) granular temperature $T_\text{s}$, the coefficient of restitution $\alpha$, the solid volume fraction $\phi$ along with the parameter $\xi_\text{b}^2$ characterizing the external energy source. In this Section we will compare our theoretical predictions for the thermostats A and B with recent Langevin dynamics simulations carried out by Gradenigo \emph{et al.} \cite{GSVP11} for hard disks ($d=2$). In these simulations, the fluid is also driven by a stochastic bath with friction and the two external parameters $\gamma_\text{b}$ and $\xi_\text{b}^2$ are related by Eq.\ \eqref{3.6} with $\beta=\frac{1}{2}$. In the steady state, they measured the static and dynamic structure factors for shear and longitudinal modes for several values of the coefficient of restitution $\alpha$ and volume fraction $\phi$. The corresponding best fit of the simulation results of the above structure factors allow them to identify the kinematic viscosity $\nu=\eta/\rho$, the longitudinal viscosity
\begin{equation}
\label{5.1bis}
\nu_l=\frac{1}{\rho}\left(2\frac{d-1}{d}\eta+\lambda\right),
\end{equation}
and the thermal diffusivity
\begin{equation}
\label{5.2bis}
D_\text{T}=\frac{2}{dn}\kappa.
\end{equation}
\begin{figure}
\includegraphics[width=0.85 \columnwidth,angle=0]{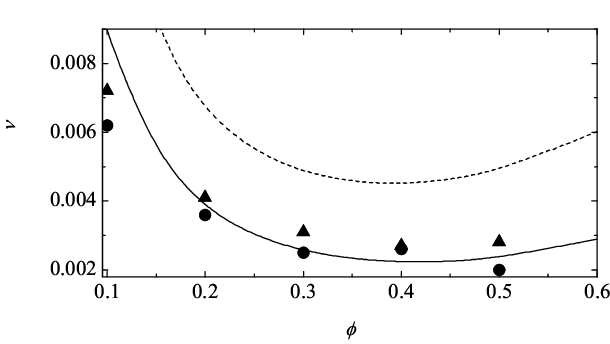}
\caption{Plot of the kinematic viscosity $\nu=\eta/\rho$ as a function of the volume fraction $\phi$ for $\alpha=0.6$. The solid line is the theoretical prediction given by Eq.\ \eqref{5.9} while the dashed line is the theoretical result obtained by assuming the elastic form of the shear viscosity $\eta$. Symbols are the simulation results obtained by Gradenigo {\em et al.} \cite{GSVP11} from the static (circles) and dynamical (triangle) correlations of transversal shear modes.
\label{fig3}}
\end{figure}
\begin{figure*}
\begin{center}
\begin{tabular}{lr}
\resizebox{7.5cm}{!}{\includegraphics{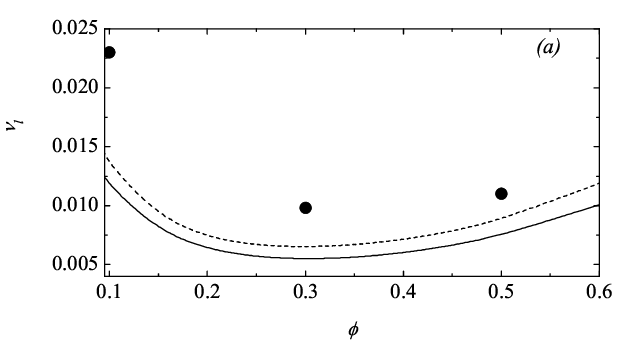}}&\resizebox{7.5cm}{!}
{\includegraphics{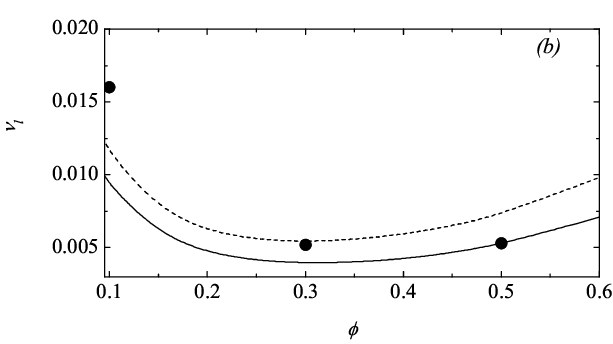}}
\end{tabular}
\end{center}
\caption{Plot of the longitudinal viscosity $\nu_l$ as a function of the volume fraction $\phi$ for two values of the coefficient of restitution: $\alpha=0.8$ (panel a), and $\alpha=0.6$ (panel b). The solid lines are the theoretical predictions for $\nu_l$ obtained by using Eqs.\ \eqref{5.7} and \eqref{5.9}  while the dashed lines are the theoretical results obtained by assuming the elastic forms of the shear viscosity $\eta$ and the bulk viscosity $\lambda$. Symbols are the simulation results obtained by Gradenigo {\em et al.} \cite{GSVP11} by fitting their numerical data for the dynamical correlations of the longitudinal modes.
\label{fig4}}
\end{figure*}
\begin{figure*}
\begin{center}
\begin{tabular}{lr}
\resizebox{7.5cm}{!}{\includegraphics{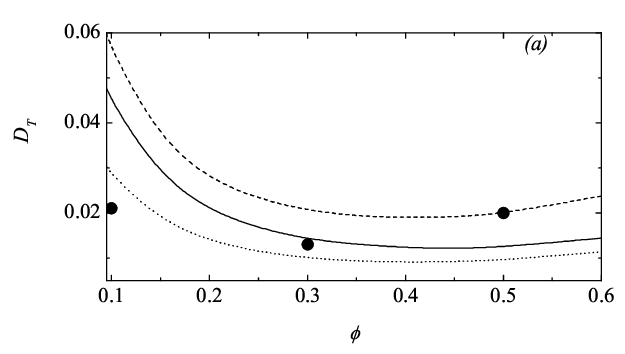}}&\resizebox{7.5cm}{!}
{\includegraphics{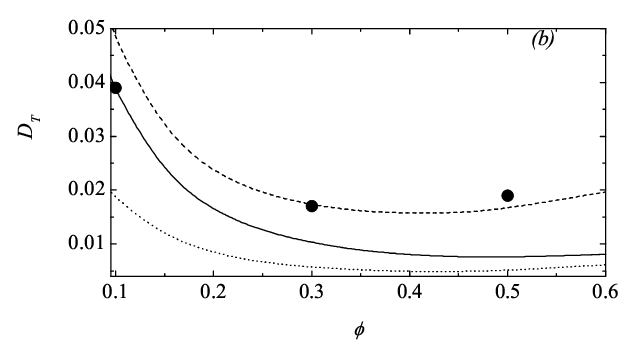}}
\end{tabular}
\end{center}
\caption{Plot of the thermal diffusivity $D_T=2\kappa/dn$ as a function of the volume fraction $\phi$ for two values of the coefficient of restitution: $\alpha=0.8$ (panel a), and $\alpha=0.6$ (panel b). Symbols are the simulation results obtained by Gradenigo {\em et al.} \cite{GSVP11} by fitting their numerical data for the dynamical correlations of the longitudinal modes. The solid lines are the theoretical predictions for $D_T$ obtained by using Eqs.\ \eqref{5.12}--\eqref{5.14}, the dotted lines are the theoretical predictions for $D_T$ obtained by using Eq.\ \eqref{5.24} and the dashed lines are the theoretical results obtained by assuming the elastic form of the thermal conductivity $\kappa$.
\label{fig5}}
\end{figure*}

Figure \ref{fig3} shows the kinematic viscosity $\nu$ for disks as a function of the volume fraction $\phi$ for $\alpha=0.6$. Symbols refer to the values of $\nu$ obtained from Langevin dynamics simulations \cite{GSVP11} by using two different procedures: (i) via the equal-time correlation of the transversal shear mode (static correlations) and (ii) via the correlation of the transversal shear mode at different times (dynamical correlations). As in Fig.\ \ref{fig1}, the parameters of the simulation are $\gamma_b=1$, $T_\text{b}=1$, $m=1$ and $\sigma=0.01$. We observe first that the simulation data obtained with the two independent procedures are compatible. Regarding the theoretical results, note that for the kinematic viscosity the results obtained by using both kind thermostats are the same. The theoretical prediction for $\eta$ in the \emph{elastic} limit (i.e., Eq.\ \eqref{5.9} with $\alpha=1$ and $\gamma_\text{b}=\xi_\text{b}=0$) but considering the $\alpha$-dependence of the granular temperature given by Eq.\ \eqref{32} is also plotted. This was the theoretical expression for $\nu$ used in Ref.\ \cite{GSVP11} to compare with simulation data. At a qualitative level, we observe that both theories (the elastic Enskog theory and the one derived here) reproduce in general the trends of simulation data. However, at a more quantitative level, it is quite apparent that the analytical results obtained in this paper for granular fluids agree much better with simulation data than those obtained in the elastic case, since the latter clearly overestimates the value of $\nu$. This is the expected result since the simulations were carried out for inelastic gases in the presence of a stochastic bath with friction.
\begin{figure}
\includegraphics[width=0.8 \columnwidth,angle=0]{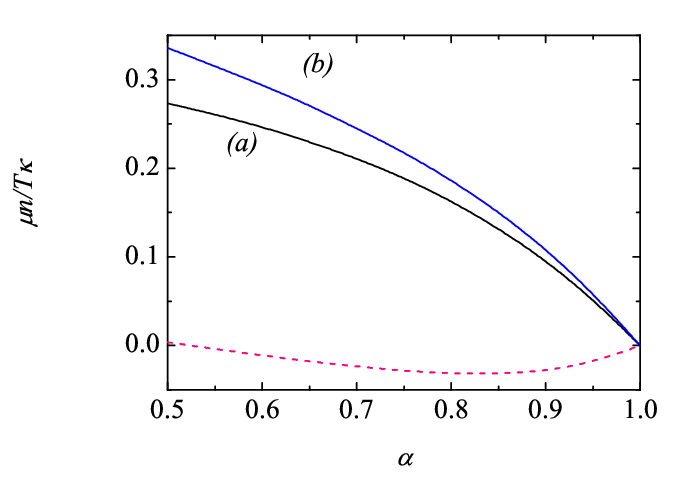}
\caption{(color online) Plot of the of the dimensionless quantity $n\mu/T\kappa$ versus the coefficient of restitution $\alpha$ for hard disks ($d=2$) with $m=1$, $\sigma=0.01$, $\gamma_\text{b}=T_\text{b}=1$ and two different values of the solid volume fraction $\phi$: (a) $\phi=0.1$, and
(b) $\phi=0.3$. The dashed line corresponds to the results obtained by considering the choice B for $\phi=0.1$. Note that $\mu=0$ in the elastic case ($\alpha=1$).
\label{fig6}}
\end{figure}
\begin{figure}
\includegraphics[width=0.75 \columnwidth,angle=0]{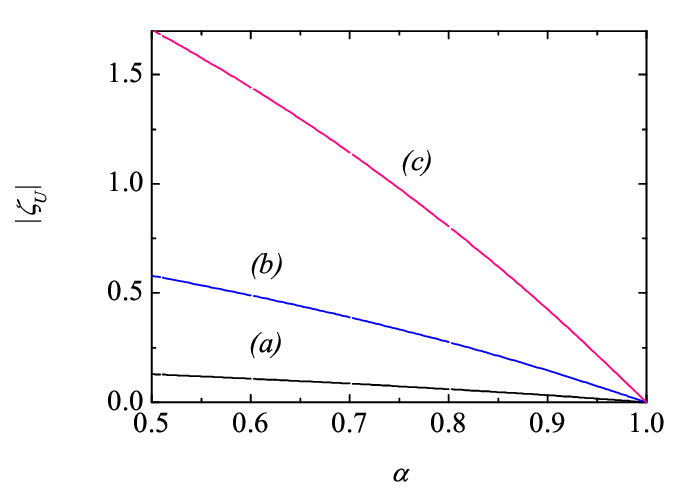}
\caption{(color online) Plot of the magnitude of the first-order contribution $\zeta_U$ to the cooling rate versus the coefficient of restitution $\alpha$ for for hard disks ($d=2$) with with $m=1$, $\sigma=0.01$, $\gamma_\text{b}=T_\text{b}=1$ and three different values of the solid volume fraction $\phi$: (a) $\phi=0.1$,
(b) $\phi=0.3$, and (c) $\phi=0.5$. Note that $\zeta_U=0$ in the elastic case ($\alpha=1$).
\label{fig7}}
\end{figure}

The longitudinal viscosity $\nu_l$ is plotted in Fig.\ \ref{fig4} versus the volume fraction $\phi$ for the same systems as in Fig.\ \ref{fig3}. We observe that, in general, the influence of the thermostat on the longitudinal viscosity is less significant than for the kinematic viscosity $\nu$ since both theories agree relatively well. However, the discrepancies with computer simulations are more important than in the case of $\nu$, specially in the low-density limit ($\phi=0.1$). While the elastic theory is closer to the simulation data than the inelastic theory when $\alpha=0.8$ (panel (a) of Fig.\ \ref{fig4}), the opposite happens at $\alpha=0.6$ for denser systems (see the panel (b) of Fig.\ \ref{fig4}). Since the dependence of the shear viscosity $\eta$ on $\phi$ is well captured by the inelastic Enskog theory (see Fig.\ \ref{fig3}), it is evident that the discrepancies between theory and simulations are essentially due to the bulk viscosity $\lambda$, whose value is specially underestimated at low-density. This is a quite surprising result since one would expect that the influence of $\lambda$ on the value of $\nu_l$ increases with increasing density since $\lambda=0$ for a dilute gas ($\phi=0$).

The thermal diffusivity is shown in Fig.\ \ref{fig5} for the same cases as those considered in Figs.\ \ref{fig3} and \ref{fig4}. Surprisingly, for strong dissipation and quite dense systems (see the panel (b) of Fig.\ \ref{fig5}), the comparison between theory and simulation agrees in general better when one uses the elastic form for $D_\text{T}$ instead of its inelastic expression \eqref{5.9}. These results contrast with the ones recently obtained \cite{G11} for the stochastic driving (i.e., when $\gamma_\text{b}\to 0$, keeping $\gamma_\text{b} T_\text{b}$ finite) where it was shown the accuracy of the inelastic Enskog theory (see Fig.\ 1 of Ref.\ \cite{G11}) for moderate densities and finite collisional dissipation. It is important to note that the identification of the transport coefficients from Langevin dynamics simulations requires to fit the simulation results for small but not zero values of the wave number $k$. Given that the expressions for the Enskog transport coefficients are independent of the wave number (since the hydrodynamic regime only strictly holds in the limit $k\to 0$), it is possible that the transport coefficients measured in the simulations are still functions of $k$, specially when the smallest value of $k$ considered to get the fit results is not close to 0. In particular, the simulation data for $\phi=0.3$ and 0.5 in the panel (b) of Fig.\ \ref{fig4} were obtained for $k\sigma=0.4$ and 0.5, respectively. In this sense, if one would extrapolate the data shown in Table 3 of Ref. \ \cite{GSVP11}, one could conclude that the true value of $D_\text{T}$ would be smaller than the one shown in this figure when $k\sigma=0$. More simulations would be needed to clarify this point.

Now we consider the $\alpha$-dependence of the transport coefficient $\mu$ and the first-order contribution $\zeta_U$ to the cooling rate. Given that both coefficients vanish in the elastic limit, they were also neglected in previous studies for heated granular fluids \cite{NETP99,GSVP11}. To assess the impact of the term $-\mu \nabla n$ in the heat flux, the reduced coefficient $\mu n/(T\kappa)$ is plotted in Fig.\ \ref{fig6} versus the coefficient of restitution for two different values of the volume fraction $\phi$ in the case of the choice A. The results derived for $\mu$ by using the choice B are also plotted for comparison in the case $\phi=0.1$. We observe that the coefficient $\mu$ is negative in the case of the choice B, although its magnitude is practically zero. This drawback ($\mu \leq 0$) of choice B is not present in the case of the choice A since $\mu$ is always positive for any value of $\alpha$ and $\phi$, similarly to what happens in the \emph{undriven} case \cite{GD99,L05}. In addition, although the magnitude of $\mu$ is in general smaller than that of the thermal conductivity $\kappa$, we observe that the influence of $\mu$ on the heat transport could not be considered negligible as the degree of dissipation increases. The $\alpha$-dependence of the magnitude of $\zeta_U$ derived from the choice A is plotted in Fig.\ \ref{fig7} for several values of the volume fraction. It is quite apparent that the influence of dissipation on $|\zeta_U|$ is more significant than in the case of $\mu$, specially at large densities. Consequently, the contribution of $\zeta_U$ to the cooling rate should be considered as the rate of dissipation increases.

\section{Linear stability analysis of the hydrodynamic equations}
\label{sec7}

The closed hydrodynamic equations for $n$, ${\bf U}$, and $T$ can be obtained by replacing the constitutive
forms of the pressure tensor, the heat flux, and the cooling rate into the balance equations
\eqref{7}--\eqref{9}. They are given by
\begin{equation}
\label{75}
D_tn+n\nabla \cdot {\bf U}=0,
\end{equation}
\begin{eqnarray}
\label{76}
D_t U_i&+&\rho^{-1}\partial_i p=\rho^{-1}\partial_j\left[\eta \left(\partial_iU_j+\partial_jU_i \right.\right.\nonumber\\
& &\left.\left.-\frac{2}{3}\delta_{ij}\nabla \cdot {\bf U}\right)+\lambda \delta_{ij}\nabla \cdot {\bf U}\right],
\end{eqnarray}
\begin{eqnarray}
\label{77}
& &\left(D_t+\frac{2\gamma_\text{b}}{m}-\frac{m\xi_\text{b}^2}{T}
+\zeta^{(0)}\right)T+\frac{2}{d n}p\nabla \cdot {\bf U}=\nonumber\\
& & \frac{2}{d n}\nabla \cdot \left
(\kappa \nabla T+\mu \nabla n\right)+\frac{2}{d n}\left[\eta \left(\partial_iU_j+\partial_jU_i \right.\right.
\nonumber\\
& & \left. \left.-\frac{2}{d}
\delta_{ij}\nabla \cdot {\bf U}\right)
+\lambda \delta_{ij}\nabla \cdot {\bf U}\right]\partial_iU_j
-T\zeta_U  \nabla\cdot {\bf U}.
\end{eqnarray}
Note that consistency would require to consider up to second order in the gradients in the expression (\ref{5.18}) for the cooling rate, since this is the order of the terms in Eqs.\ \eqref{5.6} and (\ref{5.11}) coming from the pressure tensor and the heat flux, respectively.
However, it has been shown for a dilute gas that the contributions from the cooling rate
of second order are negligible \cite{BDKS98} as compared with the corresponding contributions from Eqs.\ \eqref{5.6} and (\ref{5.11}). It is assumed here that the same holds in the dense case \cite{G05}.

The form of the Navier-Stokes equations \eqref{75}--\eqref{77} for a driven granular fluid is analogous to that of an ordinary fluid, except for the presence of the external bath parameters $\gamma_\text{b}$ and $\xi_\text{b}^2$, the contributions to the cooling rate $\zeta^{(0)}$ and $\zeta_U$ and the new transport coefficient $\mu$ in the energy balance equation. In addition, as shown in Sec.\ \ref{sec5} and depending on the values of the coefficient of restitution $\alpha$, the transport coefficients are in general different from those obtained for elastic collisions.

Equations \eqref{75}--\eqref{77} can be linearized around the stationary homogeneous state, where the hydrodynamic fields take the steady values $n_\text{s}\equiv \text{const.}$, $T_\text{s}\equiv \text{const.}$  and ${\bf U}_\text{s}={\bf 0}$.  A linear stability analysis of the hydrodynamic equations \eqref{75}--\eqref{77} have also been carried out in Ref.\ \cite{GSVP11} but neglecting any dependence of the transport coefficients on inelasticity and assuming that $\mu=\zeta_U=0$. As mentioned in the Introduction, the only impact of inelasticity on their hydrodynamic equations \cite{GSVP11} is through the $\alpha$-dependence of the (steady) granular temperature $T_\text{s}$ (see Eq.\ \eqref{32} with $a_{2,\text{s}}=0$). Thus, it is worth to assess to what extent the previous theoretical results \cite{GSVP11} are indicative of what happens when the correct expressions for the transport coefficients and the cooling rate are considered. This is the main motivation of this Section.

We assume that the deviations $\delta y_{\alpha}({\bf r},t)=y_{\alpha}({\bf r},t)-y_{\text{s} \alpha}(t)$ are small, where $\delta y_{\alpha}({\bf r},t)$ denotes the deviations of $n$, $\mathbf{U}$, and $T$ from their values in the \emph{steady} homogeneous state. To recover previous linear stability results \cite{G05} derived in the undriven case, let us consider the following (reduced) time and space variables:
\begin{equation}
\label{78}
\tau=\frac{1}{2}n_\text{s}\sigma^{d-1}\sqrt{\frac{T_\text{s}}{m}}t, \quad \mathbf{r}'=
\frac{1}{2}n_\text{s}\sigma^{d-1}{\bf r}.
\end{equation}
The dimensionless time scale $\tau$ is a measure of the average number of collisions per particle
in the time interval between $0$ and $t$. The unit length introduced in
the second equality of (\ref{78}) corresponds to the mean free path of gas
particles.

A set of Fourier transformed dimensionless variables are then
introduced by
\begin{equation}
\label{79}
\rho_{{\bf k}}(\tau)=\frac{\delta n_{{\bf k}}(\tau)}{n_{\text{s}}}, \quad
{\bf w}_{{\bf k}}(\tau)=\frac{\delta {\bf U}_{{\bf
k}}(\tau)}{\sqrt{T_\text{s}/m}},\quad \theta_{{\bf k}}(\tau)=\frac{\delta
T_{{\bf k}}(\tau)}{T_{\text{s}}},
\end{equation}
where $\delta y_{{\bf k}\alpha}\equiv \{\rho_{{\bf k}},{\bf
w}_{{\bf k}}(\tau), \theta_{{\bf k}}(\tau)\}$ is defined as
\begin{equation}
\label{80}
\delta y_{{\bf k}\alpha}(\tau)=\int d \mathbf{r}'\;
e^{-i{\bf k}\cdot \mathbf{r}'}\delta y_{\alpha}
(\mathbf{r}',\tau).
\end{equation}
Note that in Eq.\ (\ref{80}) the wave vector ${\bf k}$ is dimensionless.

In Fourier space, as expected, Eq.\ \eqref{76} shows that the $d-1$ transverse velocity components
${\bf w}_{{\bf k}\perp}={\bf w}_{{\bf k}}-({\bf w}_{{\bf k}}\cdot
\widehat{{\bf k}})\widehat{{\bf k}}$ (orthogonal to the wave vector ${\bf k}$)
decouple from the other three modes and hence can be obtained more
easily. Their evolution equation can be written as
\begin{equation}
\label{81}
\left(\frac{\partial}{\partial \tau}+\frac{1}{2}\eta^*
k^2\right){\bf w}_{{\bf k}\perp}=0,
\end{equation}
where
\begin{equation}
\label{40.1}
\eta^*=\frac{\eta}{\sigma^{1-d}\sqrt{m T_\text{s}}}
\end{equation}
The solution to Eq.\ (\ref{81}) is
\begin{equation}
\label{82}
{\bf w}_{{\bf k}\perp}({\bf k}, \tau)={\bf w}_{{\bf k}\perp}(0)\exp\left[\Lambda_\perp(k)\tau\right],
\end{equation}
where
\begin{equation}
\label{82.1}
\Lambda_\perp(k)=-\frac{1}{2}\eta^*k^2.
\end{equation}
Since the (reduced) shear viscosity coefficient $\eta^*$ is positive, then $\Lambda_\perp(k)$ becomes negative for any finite wave number $k$ and so the transversal shear modes of the driven gas are linearly stable. This result contrasts with the ones obtained in the undriven case \cite{G05} where it was shown that the transversal shear modes become \emph{unstable} for values of $k$ smaller than a certain critical wave number.

The remaining (longitudinal) modes correspond to $\rho_{{\bf k}}$, $\theta_{{\bf k}}$, and
the longitudinal velocity component of the velocity field, $w_{{\bf k}||}={\bf w}_{{\bf
k}}\cdot \widehat{{\bf k}}$ (parallel to ${\bf k}$). These modes are coupled and obey the equation
\begin{equation}
\frac{\partial \delta y_{{\bf k}\alpha }(\tau )}{\partial \tau }+M_{\alpha \beta}
\delta y_{{\bf k}\beta }(\tau)=0,
\label{83}
\end{equation}
where $\delta y_{{\bf k}\alpha }(\tau )$ denotes now the set  $\left\{\rho _{{\bf k}},\theta _{{\bf k}},
 w_{{\bf k}||}\right\}$ and $\mathsf{M}$ is the square matrix
\begin{widetext}
\begin{equation}
{\sf M}=\left(
\begin{array}{ccc}
0 & 0 & i k \\
2\sqrt{2}\zeta_0^*g+\mu^*k^2&\sqrt{2}(\zeta_0^*+2\xi^*)+D_\text{T}^*k^2 & \frac{2}{d}ik(p^*+\frac{d}{2}\zeta_U)\\
ikp^*C_\rho & ikp^* &\nu_l^*k^2
\end{array}
\right),   \label{84}
\end{equation}
\end{widetext}
where
\begin{equation}
\label{84.0}
\zeta_0^*=\frac{\ell \zeta_\text{s}^{(0)}}{\sqrt{2T_\text{s}/m}}, \quad
\xi^*=\frac{m\ell \xi_\text{b}^2}{T_\text{s}\sqrt{2T_\text{s}/m}}
\end{equation}
\begin{equation}
\label{84.1}
p^*=\frac{p_\text{s}}{n_\text{s}T_\text{s}}=1+2^{d-2}(1+\alpha)\chi \phi,
\end{equation}
and
\begin{equation}
\label{84.2}
\nu_l^*= \frac{\rho_\text{s}\nu_{l}}{2\sigma^{1-d}\sqrt{mT_\text{s}}}, \quad
D_\text{T}^*= \frac{n_\text{s}D_\text{T}}{2\sigma^{1-d}\sqrt{T_\text{s}/m}},
\end{equation}
\begin{equation}
\label{84.2.1}
\mu^*=\frac{\rho_\text{s}}{d\sigma^{1-d}T_\text{s}\sqrt{mT_\text{s}}}\mu.
\end{equation}
Here, $\rho_\text{s}=m n_\text{s}$ is the mass density. In the above equations, it is understood that the transport coefficients $\eta$, $\nu_l$, $D_\text{T}$, and $\mu$ are evaluated in the homogeneous steady state. In addition, the quantities $g(\phi)$ and $C_\rho(\alpha,\phi)$ appearing in the matrix ${\sf M}$ are given by
\begin{equation}
\label{85}
g(\phi)=1+\phi\frac{\partial}{\partial \phi}\ln \chi(\phi),
\end{equation}
\begin{eqnarray}
\label{86}
C_\rho(\alpha,\phi)&=&1+\phi\frac{\partial}{\partial \phi}\ln p^*(\alpha,\phi)
\nonumber\\
&=&1+g(\phi)-\frac{g(\phi)}{1+2^{d-2}(1+\alpha)\phi \chi(\phi)},\nonumber\\
\end{eqnarray}
where in the last equality use has been made of the explicit expression of $p^*$
given by Eq.\ (\ref{84.1}). If one assumes $\mu^*=\zeta_U=0$, the matrix \eqref{84} agrees with the dynamical matrix obtained when the gas is heated by a stochastic thermostat ($\gamma_\text{b}=0$ but $\gamma_\text{b} T_\text{b}=\text{finite}$ and $\zeta_0^*=\xi^*$) \cite{NETP99}.

The longitudinal three modes have the form $\exp[\Lambda_\ell(k) \tau]$ for $\ell=1,2,3$, where
$\Lambda_\ell(k)$ are the eigenvalues of the matrix ${\sf M}$, namely, they are the solutions of the cubic equation
\begin{equation}
\label{86.1}
\Lambda^3+A(k) \Lambda^2+B(k) \Lambda+C(k)=0,
\end{equation}
where
\begin{equation}
\label{87}
A(k)=\sqrt{2}(\zeta_0^*+2\xi^*)+k^2\left(\nu_l^*+D_\text{T}^*\right),
\end{equation}
\begin{eqnarray}
\label{88}
B(k)&=&k^4\nu_l^*D_\text{T}^*
+k^2\left[p^*C_\rho+p^*\left(\frac{2}{d}p^*+\zeta_U\right)\right.\nonumber\\
& & \left.
+\sqrt{2}(\zeta_0^*+2\xi^*)\nu_l^*\right],
\end{eqnarray}
\begin{eqnarray}
\label{89}
C(k)&=&p^*k^2\left[\sqrt{2}C_\rho\left(\zeta_0^*+2\xi^*\right)
-2\sqrt{2}g\zeta_0^*\right.\nonumber\\
& & \left.+\left(C_\rho D_\text{T}^*-\mu^*\right)k^2\right].
\end{eqnarray}
One of the longitudinal modes (the heat mode) could be unstable for $k<k_\text{h}$, where $k_\text{h}$ is obtained from Eq.\ \eqref{86.1} when $\Lambda=0$, namely, $C(k_h)=0$. The result is
\begin{equation}
\label{90}
k_\text{h}^2=\sqrt{2}\frac{2g\zeta_0^*-C_\rho(\zeta_0^*+2\xi^*)}{C_\rho D_\text{T}^*-\mu^*}.
\end{equation}
On the other hand, an analysis of the dependence of $k_\text{h}^2$ on the coefficient of restitution $\alpha$ and the volume fraction $\phi$ shows that  $k_\text{h}^2<0$ for any value of $\alpha$ and $\phi$. Thus,
there are no physical values of $k_\text{h}$ for which the heat mode becomes unstable. Consequently, \emph{all} the eigenvalues of the dynamical matrix ${\sf M}$ have a \emph{positive} real part and no instabilities are found due to the presence of the external bath. This conclusion agrees with the results obtained in Refs.\ \cite{NETP99} and \cite{GSVP11} for driven granular fluids.

In summary, the results obtained here including the complete $\alpha$-dependence of the transport coefficients show no new surprises relative to the earlier works \cite{NETP99,GSVP11} by considering the elastic Enskog expressions for the above coefficients. Of course, the quantitative forms for the dispersion relations can be quite different in both (elastic and inelastic) approaches since the impact of dissipation on the transport coefficients and the cooling rate is significant and so, their functional forms differ appreciably from their elastic forms.


\section{Discussion}
\label{sec8}

In this paper, we have determined the transport coefficients of a granular fluid driven by a stochastic bath with friction. The results have been obtained within the framework of the (inelastic) Enskog kinetic theory and they are expected to apply over a wide range of densities. Our goal is not only academic since, from a practical standpoint, many of the simulations reported \cite{thermostat,NETP99,VAZ11,GSVP11} for flowing granular materials have used external driving forces to fluidize the system. For this reason, it would be convenient to provide to simulators with the corresponding expressions of the transport coefficients when the granular fluid is heated by a thermostat. In fact, due to the lack of the above expressions, in most of the cases it is assumed that the forms of the transport coefficients of the driven granular fluid are the same as those given by the \emph{elastic} Enskog theory \cite{FK72}. However, as expected from previous theoretical works \cite{GM02,G11}, the present results show again that the expressions for the transport coefficients clearly differ from those obtained for ordinary fluids so that, one should use the true inelastic Enskog coefficients to analyze granular flows driven by thermostats.

The transport processes considered are those for a driven fluid with small spatial gradients of the hydrodynamic fields. In this situation, the Enskog equation has been solved by means of the Chapman-Enskog method \cite{CC70} up to the first order in the spatial gradients. Since these gradients have been assumed to be independent of the coefficient of restitution, although the corresponding hydrodynamic equations restrict their applicability to first order in gradients, the transport coefficients appearing in these equations hold a \emph{priori} to arbitrary degree of dissipation.

An important but subtle point is the generalization of the driving external forces (which are mainly used in homogeneous situations) to \emph{inhomogeneous} states. This is a crucial step since one has to consider situations close to steady homogeneous states to determine the transport coefficients from the Chapman-Enskog expansion. Although the above generalization is a matter of choice, it has important implications in the final expressions of the transport coefficients. For simplicity, in previous works on heated granular gases \cite{GM02,G11} it was assumed that the external driving force has the same form as in the homogeneous case, except that their parameters are local quantities. As a consequence, the parameters of the force are chosen to impose a stationary temperature in the zeroth-order solution (i.e., $\partial_t^{(0)}T=0$). However, for general small perturbations around the steady homogeneous state, it is expected that the density and temperature are specified separately in the local reference state $f^{(0)}$ and so, the temperature cannot be stationary at any point of the system (i.e., $\partial_t^{(0)}T\neq 0$). This choice is more general than the previous one ($\partial_t^{(0)}T=0$) and has the advantage of a simpler implementation on computer simulations since the parameters of the driven external force are constant, even for inhomogeneous states.

As mentioned in the Introduction, the fact that $\partial_t^{(0)}T\neq 0$ gives rise to conceptual and practical difficulties not present in the case of the choice B. One of them is that the evaluation of the complete nonlinear dependence of the transport coefficients on dissipation requires in principle the analysis of the hydrodynamic behavior of the \emph{unsteady} reference state. This involves the corresponding numerical integration of the differential equations obeying the velocity moments of the zeroth-order distribution  $f^{(0)}$ (see for instance, Eq.\ \eqref{3.8} for the fourth degree moment $a_2$ of $f^{(0)}$). This is quite an intricate long problem. However, given that here we are interested in evaluating the momentum and heat fluxes in the first order of the deviations from the steady reference state, the transport coefficients must be determined to zeroth order in the deviations. As a consequence, the steady-state condition \eqref{20.1} applies and the transport coefficients and the cooling rate can be defined in terms of the hydrodynamic fields in the steady state. Explicit expressions for these quantities have been obtained after considering the leading terms in a Sonine polynomial expansion. These explicit forms have been displayed in Sec.\ \ref{sec5} and Appendix \ref{appD} for the choices A and B, respectively. More specifically, in the case of the choice A, the bulk $\lambda$ and shear $\eta$ viscosities  are given by Eqs.\ \eqref{5.7} and \eqref{5.9}, respectively, the thermal conductivity $\kappa$ is given by Eqs.\ \eqref{5.12} and \eqref{5.14}, the coefficient $\mu$ is given by Eqs.\ \eqref{5.16} and \eqref{5.17} and the cooling rate $\zeta$ is defined by Eqs.\ \eqref{5.18}--\eqref{5.23}. All these expressions clearly show the complex dependence of the set $\left\{\lambda,\eta, \kappa, \mu, \zeta\right\}$ on the granular temperature $T$, the coefficient of restitution $\alpha$, the solid volume fraction $\phi$ and the model parameter $\xi_\text{b}^2$. In the case of the choice B, our results show that the expressions of $\lambda$ and $\eta$ are the same as those obtained from the choice A but the forms of $\kappa$ and $\mu$ are different (they are given by Eqs.\ \eqref{5.24} and \eqref{5.25}, respectively). An important drawback of the results derived from the choice B is that the coefficient $\mu$ can be negative (see Fig.\ \ref{fig6}), although its magnitude is very small.

A comparison with recent Langevin dynamics simulations \cite{GSVP11} carried out for a granular fluid driven also by a stochastic bath with friction has been made in Sec.\ \ref{sec6}. The comparison has been displayed in Fig.\ \ref{fig3} for the kinematic viscosity $\nu$, Fig.\ \ref{fig4} for the longitudinal viscosity $\nu_l$ and Fig.\ \ref{fig5} for the thermal diffusivity $D_T$. It is quite apparent that while the predictions of the driven kinetic theory compares very well with simulation data for $\nu$ in a wide range of densities, some discrepancies appear in the cases of $\nu_l$ and $D_T$ as the gas becomes denser. Surprisingly, in the case of $D_T$, the comparison agrees better when one uses the elastic form of $D_T$ in the more inelastic system ($\alpha=0.6$) studied. We think that this disagreement is in part due to the fact that while the simulation data have been obtained for small but \emph{finite} values of the wave number $k$, the Enskog expressions for the transport coefficients only strictly apply in the limit $k\to 0$. Moreover, given that these discrepancies appear at sufficiently high densities, it could also reflect the limitations of the Enskog equation (which is based on the molecular chaos hypothesis) as the granular fluid becomes denser.

With these new expressions for the momentum and heat fluxes and the cooling rate, a closed set of hydrodynamic equations for situations close to homogeneous steady states has been derived. A stability analysis of these linearized hydrodynamic equations with respect to the homogeneous steady state has been carried out to identify the conditions for stability in terms of dissipation. Our results show that the driven homogeneous state is \emph{stable} for any value of dissipation at sufficiently long wavelengths. This conclusion agrees with previous findings \cite{NETP99,GSVP11} obtained by using the \emph{elastic} expressions of the transport coefficients.

An interesting point is the usefulness of the theoretical results derived in this paper to modelize the experiments performed by using boundary driven conditions. As usual in computer simulations \cite{thermostat}, in this paper we have fluidized the system by means of a thermostat composed by a friction term which mimics the presence of an interstitial fluid and a stochastic force that models the effect of a vibrating wall. The main advantage of using this type of driving mechanism is the possibility of making theoretical progress. In addition, although the relationship of the last external force with real vibrating walls is not clear to date, some theoretical results (see, for instance Fig.\ 2 of Ref.\ \cite{G09}) obtained for the temperature ratio of a granular impurity immersed in a granular gas heated by the stochastic thermostat compare quite well with molecular dynamics simulations of shaken mixtures \cite{SUKSS06}. This agreement could stimulate the use of this simple stochastic driving for qualitative comparisons with experimental results. On the other hand, more comparisons between kinetic theory results for heated granular gases and computer simulations performed in realistic vibrating beds are needed before qualitative conclusions can be drawn.

Finally, an extension of the results derived in this paper to the more realistic shear flow problem could be an interesting project for the next future. Another possible future work is the extension of the present results to the important subject of granular mixtures. Given the difficulties associated with multicomponent systems, the tracer diffusion could be perhaps a good starting point to provide some insight into the general problem. Work along the above lines will be carried out in the near future.

\acknowledgments

The authors are grateful to Maribel Garc\'{\i}a de Soria and Pablo Maynar for valuable discussions and for sending a preprint of their unpublished results. The present work has been supported by the Ministerio de
Educaci\'on y Ciencia (Spain) through grants No. FIS2010-16587 (V.G., M.G.Ch. and F.V.) and No. MAT2009-14351-C02-02 (F.V.). The first Grant has been partially financed by
FEDER funds and by the Junta de Extremadura (Spain) through Grant No. GRU10158. The  research  of M. G. Chamorro  has  been  supported  by  the  predoctoral  fellowship BES-2011-045869 from the  Spanish Government (Spain).

\appendix
\section{Behavior of the fourth-cumulant of the zeroth-order distribution in the vicinity of the steady state}
\label{appA}

Although the determination of $a_2(\xi^*)$ requires numerical work, one can obtain analytically this quantity in the vicinity of the steady state by means of the derivative $\partial a_2/\partial \xi^*$ evaluated at the steady state. This derivative appears in the expressions of the thermal conductivity $\kappa$ (see Eq.\ \eqref{5.14}) and the first-order contribution $\zeta_U$ to the cooling rate (see Eq.\ \eqref{5.21}). This Appendix addresses the evaluation of the above derivative.

In order to determine $\partial a_2/\partial \xi^*$ from Eq.\ \eqref{3.8}, we first assume that $\varphi$ can be well described by the lowest Sonine approximation \eqref{29.1}. Then, approximate forms for $\zeta_0^*=(2/d)\mu_2$ and $\mu_4$ can be obtained when one uses the distribution \eqref{29.1} and neglects nonlinear terms in $a_2$. The results are \cite{NE98}
\begin{equation}
\label{a2} \mu_2\to \mu_2^{(0)}+\mu_2^{(1)}a_2,\quad \mu_4\to
\mu_4^{(0)}+\mu_4^{(1)}a_2,
\end{equation}
where
\begin{equation}
\label{a3}
\mu_2^{(0)}=\frac{\pi^{(d-1)/2}}{\sqrt{2}\Gamma\left(\frac{d}{2}\right)}\chi (1-\alpha^2),\quad
\mu_2^{(1)}=\frac{3}{16}\mu_2^{(0)},
\end{equation}
\begin{equation}
\label{a4} \mu_4^{(0)}=\left(d+\frac{3}{2}+\alpha^2\right)\mu_2^{(0)},
\end{equation}
\begin{equation}
\label{a5}
\mu_4^{(1)}=\left[\frac{3}{32}(10d+39+10\alpha^2)+\frac{d-1}{1-\alpha}\right]\mu_2^{(0)}.
\end{equation}
With the use of the approximations \eqref{a2} and retaining only linear terms in $a_2$ in Eq.\ \eqref{3.8}, the derivative $\partial a_2/\partial \xi^*$ is given by
\begin{widetext}
\begin{equation}
\label{a6}
\frac{\partial a_2}{\partial \xi^*}=-\frac{\mu_4^{(0)}-(d+2)\mu_2^{(0)}-\left[\frac{19}{16}(d+2)
\mu_2^{(0)}-\mu_4^{(1)}-\frac{d(d+2)}{2}\xi^*\right]a_2}{\frac{3d(d+2)}{8}\xi^*\left[(2\beta T^*-1)\xi^*+\frac{2}{d}\left(\mu_2^{(0)}+\mu_2^{(1)}a_2\right)\right]}.
\end{equation}
\end{widetext}
However, some care must be taken in Eq.\ \eqref{a6} at the steady state since the numerator and denominator of Eq.\ \eqref{a6} vanish and so, the corresponding expression for the derivative $\partial a_2/\partial \xi^*$ becomes indeterminate. This difficulty can be solved by means of l'Hopital's rule. After some algebra, it is straightforward to see that the steady-state value of the derivative $\Delta\equiv (\partial a_2/\partial \xi^*)_\text{s}$ obeys the quadratic equation
\begin{eqnarray}
\label{a7}
&&\frac{3}{4}(d+2)\mu_2^{(1)}\xi_\text{s}^*\Delta^2+\left[\frac{d(d+2)}{8}\left(1+2\beta T_\text{s}^*\right)\xi_\text{s}^*
\right.\nonumber\\
& & \left.-\frac{19}{16}(d+2)\mu_2^{(0)}+\mu_4^{(1)}\right]
\Delta+\frac{d(d+2)}{2}a_{2,\text{s}}=0,
\end{eqnarray}
where $T_\text{s}^*=T_\text{s}/T_\text{b}$. Since $a_{2,\text{s}}$ is in general very small, it is expected that the magnitude of $\Delta$ be also quite small. An analysis of the solutions to Eq.\ \eqref{a7} shows that in general one of its roots is much larger than $a_{2,\text{s}}$ while the other is of the order of $a_{2,\text{s}}$. We take the latter one as the physical root of the quadratic equation \eqref{a7}.

\vicente{Since $\Delta$ is in general very small, one may neglect the term proportional to $\Delta^2$ in Eq.\ \eqref{a7}. In this case, the derivative $\Delta$ can be explicitly written as
\begin{equation}
\label{a8}
\Delta=\frac{a_{2,\text{s}}}{\frac{19}{8d}\mu_2^{(0)}-\frac{1+2\beta T_\text{s}^*}{4}\xi_\text{s}^*-\frac{2}{d(d+2)}\mu_4^{(1)}}.
\end{equation}
The derivative $(\partial a_2/\partial \theta)_\text{s}$ can be directly derived from Eq.\ \eqref{a6} with the result
\begin{equation}
\label{a9}
\left(\frac{\partial a_2}{\partial \theta}\right)_\text{s}=\frac{\xi_\text{s}^{*4/3}\Delta}
{\frac{19}{12d}\mu_2^{(0)}-\frac{2}{3}\xi_\text{s}^*-\frac{4}{3d(d+2)}\mu_4^{(1)}-\frac{\mu_{2}^{(1)}}{d}\xi_\text{s}^*\Delta}.
\end{equation}}

\section{First order approximation}
\label{appB}

The application of the Chapman-Enskog method up to the first order approximation follows similar mathematical steps as those made before in the undriven case \cite{GD99,L05,G12}. Some details on this derivation are provided in this Appendix. Up to the first order in the expansion, the velocity distribution function $f^{(1)}$ obeys the kinetic equation
\begin{eqnarray}
& &\left(\partial_{t}^{(0)}+{\cal L}\right)f^{(1)}-\frac{\gamma_\text{b}}{m}
\frac{\partial}{\partial {\bf v}}\cdot {\bf V}f^{(1)}-\frac{1}{2}\xi_\text{b}^2\frac{\partial^2}{\partial v^2}f^{(1)}
\nonumber\\
& &
=-\left(\partial_{t}^{(1)}+{\bf v}\cdot \nabla \right)f^{(0)}-J_\text{E}^{(1)}[f].\nonumber\\ \label{b1}
\end{eqnarray}
Here, $J^{(1)}_\text{E}[f]$ means the first order contribution to the expansion of the Enskog collision operator and ${\cal L}$ is the linear operator
\begin{equation}
\label{b2}
{\cal L}f^{(1)}=-\left(J_\text{E}^{(0)}[f^{(0)},f^{(1)}]+J_\text{E}^{(0)}[f^{(1)},f^{(0)}]\right).
\end{equation}
The macroscopic balance equations to first order in the gradients are
\begin{equation}
\label{b3}
D_t^{(1)}n=-n\nabla\cdot {\bf U},\quad
D_t^{(1)}U_i=-(mn)^{-1}\nabla_i p,
\end{equation}
\begin{equation}
\label{b3.1}
D_t^{(1)}T=-\frac{2p}{dn}\nabla\cdot {\bf U}-\zeta^{(1)}T,
\end{equation}
where $D_t^{(1)}\equiv \partial_t^{(1)}+{\bf U}\cdot \nabla$ and $\zeta^{(1)}$ is the first order contribution to the cooling rate. Use of Eqs.\ (\ref{b3}) in (\ref{b1}) and taking into account the form of $J_\text{E}^{(1)}[f]$ obtained in Ref.\  \cite{GD99} for the undriven case, one gets
\begin{eqnarray}
\label{b4}
& & \left(\partial_{t}^{(0)}+{\cal L}\right)f^{(1)}-\frac{\gamma_\text{b}}{m}
\frac{\partial}{\partial {\bf v}}\cdot {\bf V}
f^{(1)}-\frac{1}{2}\xi_\text{b}^2\frac{\partial^2}{\partial v^2}f^{(1)}\nonumber\\
& &
={\bf A}\cdot
\nabla \ln T+{\bf B}\cdot \nabla \ln n\nonumber\\
& & +C_{ij}\frac{1}{2}\left( \partial _{i}U_{j}+\partial _{j
}U_{i}-\frac{2}{d}\delta _{ij}\nabla \cdot\mathbf{U} \right)+D \nabla \cdot\mathbf{U},\nonumber\\
\end{eqnarray}
where
\begin{equation}
{\bf A}\left( \mathbf{V}\right)=-\mathbf{V}T\partial_T f^{(0)}
-\frac{p}{\rho }\frac{\partial f^{(0)}}{\partial \mathbf{V}}
-\boldsymbol{\mathcal{K}}[T\partial_T f^{(0)}],  \label{b5}
\end{equation}
\begin{eqnarray}
{\bf B}\left(\mathbf{V}\right)&=& -{\bf V}n\partial_nf^{(0)}-\frac{p}{\rho}
\left(1+\phi\frac{\partial}{\partial \phi}\ln p^*\right)
\frac{\partial f^{(0)}}{\partial \mathbf{V}}\nonumber\\
& &-\boldsymbol{\mathcal{K}}[n\partial_nf^{(0)}]-\frac{1}{2}\phi\frac{\partial}{\partial \phi}\ln \chi
\boldsymbol{\mathcal{K}}[f^{(0)}],  \label{b6}
\end{eqnarray}
\begin{equation}
\label{b7} C_{ij}\left(
\mathbf{V}\right)=V_i\frac{\partial}{\partial V_j}f^{(0)}+\mathcal{K}_i[\partial_{V_j} f^{(0)}],
\end{equation}
\begin{eqnarray}
D&=&\frac{1}{d}\frac{\partial}{\partial \mathbf{V}}\cdot (\mathbf{V}
f^{(0)})+\left( \zeta _{U}+\frac{2}{d} p^*\right)
T\partial_T f^{(0)}\nonumber\\
& & -f^{(0)}+n\partial_n f^{(0)}+\frac{1}{d}\mathcal{K}_{i}[\partial_{V_i} f^{(0)}].   \label{b8}
\end{eqnarray}
Here, $\phi$ and $p^*$ are given by Eqs.\ \eqref{33}
and \eqref{84.1}, respectively,
\vicente{
\begin{equation}
\label{b8.1}
n\frac{\partial f^{(0)}}{\partial n}=f^{(0)}-\xi^*\frac{\partial f^{(0)}}{\partial \xi^*}
-\frac{2}{3}\theta\frac{\partial f^{(0)}}{\partial \theta}+\phi \frac{\partial \chi}{\partial \phi}
\frac{\partial f^{(0)}}{\partial \chi},
\end{equation}}
and $\mathcal{K}_{i}$ is the operator
\begin{eqnarray}
\mathcal{K}_{i}[\mathbf{V}_1|X]&=&\sigma^{d}\chi\int d \mathbf{v}_{2}\int d\widehat{\boldsymbol {\sigma
}}\Theta (\widehat{\boldsymbol {\sigma}} \cdot
\mathbf{g}_{12})(\widehat{\boldsymbol {\sigma}}\cdot
\mathbf{g}_{12})
\widehat{\sigma}_i\nonumber\\
& \times& \left[ \alpha
^{-2}f^{(0)}(\mathbf{V}_{1}^{\prime
\prime})X(\mathbf{V}_{2}^{\prime \prime
})+f^{(0)}(\mathbf{V}_{1})X(\mathbf{V}_{2})\right], \nonumber\\ \label{b10}
\end{eqnarray}
where ${\bf V}_1^{\prime\prime}={\bf V}_1-\frac{1}{2}(1+\alpha^{-1})(\widehat{\boldsymbol {\sigma }}\cdot
\mathbf{g}_{12})\widehat{\boldsymbol {\sigma}}$, ${\bf V}_2''={\bf V}_2+\frac{1}{2}(1+\alpha^{-1})(\widehat{\boldsymbol {\sigma}}\cdot\mathbf{g}_{12})\widehat{\boldsymbol {\sigma }}$, and $\mathbf{g}_{12}=\mathbf{V}_1-\mathbf{V}_2$. In Eq.\ \eqref{b8}, $\zeta_U$ is defined through the expression
\begin{equation}
\label{b11}
\zeta=\zeta^{(0)}+\zeta_U\nabla \cdot {\bf U},
\end{equation}
where $\zeta^{(0)}$ is the cooling rate evaluated at zeroth-order. In addition, upon deriving Eqs.\ \eqref{b4}--\eqref{b8}, use has been made of the spherical symmetry of $f^{(0)}$ which allows us to write the tensor derivative of the flow field $\partial_i U_j$ in terms of its independent trace and traceless parts, e.g.,
\begin{eqnarray}
\label{b11.1}
V_i\frac{\partial f^{(0)}}{\partial V_j}\partial_i U_j&=&V_i\frac{\partial f^{(0)}}{\partial V_j}\frac{1}{2}\left(\partial_j U_i+\partial_i U_j\right)\nonumber\\
&=&V_i\frac{\partial f^{(0)}}{\partial V_j}\frac{1}{2}\left(\partial_j U_i+\partial_i U_j-\frac{2}{d}\delta_{ij}\nabla \cdot \mathbf{U}\right)\nonumber\\
& & +
\frac{1}{d}\mathbf{V}\cdot \frac{\partial f^{(0)}}{\partial \mathbf{V}}\nabla \cdot \mathbf{U},
\end{eqnarray}
and a similar analysis of the contribution from $\mathcal{K}_i[\partial_{V_j} f^{(0)}]$.

The solution to Eq.\ \eqref{b4} can be written in the form \eqref{5.1}, where $\boldsymbol{\mathcal{A}}$, $\boldsymbol{\mathcal{B}}$, $\mathcal{C}_{ij}$, and $\mathcal{D}$ are unknown functions of the peculiar velocity.   Since the gradients of the hydrodynamic fields are all independent, Eq.\ (\ref{b4}) can be separated into independent equations for each coefficient. This yields the following set of linear, inhomogeneous integral equations:
\begin{eqnarray}
\label{b13}
& & -\left(\frac{2\gamma_\text{b}}{m}-\frac{m\xi_\text{b}^2}{T} +\zeta^{(0)}\right)T\partial_T
\boldsymbol{\mathcal{A}}-\left[
\frac{m}{T}\xi_\text{b}^2\left(1-\frac{3}{2}\frac{\partial \zeta_0^*}{\partial \xi^*}\right)\right.\nonumber\\
& &\left. +\frac{1}{2}\zeta^{(0)}\right]
\boldsymbol{\mathcal{A}}-\frac{\gamma_\text{b}}{m}\frac{\partial}{\partial {\bf v}}\cdot {\bf V}\boldsymbol{\mathcal{A}}
-\frac{1}{2}\xi_\text{b}^2
\frac{\partial^2}{\partial v^2}\boldsymbol{\mathcal{A}}+{\cal L}\boldsymbol{\mathcal{A}}={\bf A},\nonumber\\
\end{eqnarray}
\vicente{
\begin{eqnarray}
\label{b14}
& &-\left(\frac{2\gamma_\text{b}}{m}-\frac{m\xi_\text{b}^2}{T} +\zeta^{(0)}\right)T\partial_T
\boldsymbol{\mathcal{B}}
-\frac{\gamma_\text{b}}{m}\frac{\partial}{\partial {\bf v}}\cdot {\bf V}\boldsymbol{\mathcal{B}}\nonumber\\
& & -\frac{1}{2}\xi_\text{b}^2
\frac{\partial^2}{\partial v^2}\boldsymbol{\mathcal{B}}+{\cal L}\boldsymbol{\mathcal{B}}={\bf B}+
\zeta ^{(0)}\left(1+\phi\frac{\partial}{\partial \phi}\ln \chi \right)\boldsymbol{\mathcal{A}}\nonumber\\
&&+\left[\phi \frac{\partial \chi}{\partial \phi}\frac{\partial}{\partial \chi}\left(\frac{\zeta^{(0)}}{\chi}\right)-\xi^*\frac{\partial \zeta^{(0)}}{\partial \xi^*}
-\frac{2}{3}\theta\frac{\partial \zeta^{(0)}}{\partial \theta}\right]
\boldsymbol{\mathcal{A}},
\end{eqnarray}}
\begin{eqnarray}
\label{b15}
& &-\left(\frac{2\gamma_\text{b}}{m}-\frac{m\xi_\text{b}^2}{T} +\zeta^{(0)}\right)T\partial_T
\mathcal{C}_{ij}
-\frac{\gamma_\text{b}}{m}\frac{\partial}{\partial {\bf v}}\cdot {\bf V}\mathcal{C}_{ij}\nonumber\\
& & -\frac{1}{2}\xi_\text{b}^2
\frac{\partial^2}{\partial v^2}\mathcal{C}_{ij}+{\cal L}\mathcal{C}_{ij}=C_{ij},  \nonumber\\
\end{eqnarray}
\begin{eqnarray}
\label{b16}
&&-\left(\frac{2\gamma_\text{b}}{m}-\frac{m\xi_\text{b}^2}{T} +\zeta^{(0)}\right)T\partial_T
\mathcal{D}
-\frac{\gamma_\text{b}}{m}\frac{\partial}{\partial {\bf v}}\cdot {\bf V}\mathcal{D}\nonumber\\
& &-\frac{1}{2}\xi_\text{b}^2
\frac{\partial^2}{\partial v^2}\mathcal{D}+{\cal L}\mathcal{D}=D,  \nonumber\\
\end{eqnarray}
Upon deriving Eqs.\ \eqref{b13}--\eqref{b16} use has been made of the result
\vicente{
\begin{eqnarray}
\partial _{t}^{(0)}\nabla \ln T&=&\nabla \partial_{t}^{(0)}\ln T=\nabla
\left(\frac{m\xi_\text{b}^2}{T}-\frac{2}{m}\gamma_\text{b} -\zeta^{(0)}\right)\nonumber\\
&=&-\zeta ^{(0)}\left(1+\phi\partial_{\phi}\ln \chi \right)
\nabla \ln n \nonumber\\
&& -\left[\phi \frac{\partial \chi}{\partial \phi}\frac{\partial}{\partial \chi}\left(\frac{\zeta^{(0)}}{\chi}\right)-\xi^*\frac{\partial \zeta^{(0)}}{\partial \xi^*}
-\frac{2}{3}\theta\frac{\partial \zeta^{(0)}}{\partial \theta}\right]
\nabla \ln n \nonumber\\
& &-\left[\frac{m}{T}\xi_\text{b}^2\left(1-\frac{3}{2}\frac{\partial \zeta_0^*}{\partial \xi^*}\right)+\frac{1}{2}\zeta ^{(0)}\right]\nabla \ln T.
\nonumber\\
\label{b17}
\end{eqnarray}}

As noted in Section \ref{sec5}, in the first order of the deviations from the steady state, we only need to know the transport coefficients to zeroth order in the deviations (steady state conditions). This means that the term
$$
\frac{2\gamma_\text{b}}{m}-\frac{m\xi_\text{b}^2}{T} +\zeta^{(0)}
$$ appearing in the left-hand side of Eqs.\ \eqref{b13}--\eqref{b16} vanishes. The differential equations for the transport coefficients thus become simple coupled algebraic equations. They are given by Eqs.\ \eqref{5.2}--\eqref{5.5}.

\section{Kinetic contributions to the transport coefficients}
\label{appC}

In this Appendix we determine from Eqs.\ \eqref{5.2}--\eqref{5.5} the kinetic contributions to the transport coefficients $\eta$, $\kappa$ and $\mu$ as well as the first order contribution $\zeta_U$ to the cooling rate. Given that all these coefficients are evaluated in the steady state, the subscript $\text{s}$ appearing along the main text will be omitted in this Appendix for the sake of brevity.

We start with the shear viscosity $\eta$. Its kinetic part $\eta_k$ is given by
\begin{equation}
\label{c1}
\eta_k=-\frac{1}{(d-1)(d+2)}\int\; d{\bf v} D_{ij} \mathcal{C}_{ij}({\bf V}),
\end{equation}
where
\begin{equation}
\label{c1.1}
D_{ij}=m\left(V_iV_j-\frac{1}{d}V^2\delta_{ij}\right).
 \end{equation}
To obtain $\eta_k$, we multiply Eq.\ (\ref{5.4}) by $D_{ij}$ and integrate over velocity. The result is
\begin{eqnarray}
\label{c2}
\left(\frac{2\gamma_\text{b}}{m}+\nu_\eta\right)\eta_k&=&nT-\frac{1}{(d-1)(d+2)}\int\; d{\bf V}
D_{ij}({\bf V}) \nonumber\\
& & \times\mathcal{K}_i\left[\frac{\partial}{\partial V_j}f^{(0)}
\right],
\end{eqnarray}
where
\begin{equation}
\label{c3}
\nu_\eta=\frac{\int d{\bf v} D_{ij}({\bf V}){\cal L}{\cal C}_{ij}({\bf V})}
{\int d{\bf v}D_{ij}({\bf V}){\cal C}_{ij}({\bf V})}.
\end{equation}
The collision integral of the right hand side of Eq.\ (\ref{c2}) has been evaluated in previous works \cite{GD99,L05}:
\begin{eqnarray}
\label{c3.1}
\int\; d{\bf V}\; D_{ij}({\bf V}) \mathcal{K}_i\left[\frac{\partial f^{(0)}}{\partial V_j}
\right]&=&2^{d-2}(d-1)nT\chi \phi \nonumber\\
&\times& (1+\alpha)(1-3\alpha).
\end{eqnarray}
Thus, the kinetic part $\eta_k$ can be written as
\begin{equation}
\label{c4}
\eta_k=\frac{nT}{\nu_\eta+\frac{2\gamma_\text{b}}{m}}
\left[1-\frac{2^{d-2}}{d+2}(1+\alpha)
(1-3 \alpha)\phi \chi \right].
\end{equation}
In order to get an explicit expression for $\eta_k$, one has to evaluate the (reduced) collision frequency $\nu_\eta^*$. It can be evaluated by considering the leading terms in a Sonine polynomial expansion of the function $\mathcal{C}_{ij}({\bf V})$. Here, we have considered a recent modified version of the standard method \cite{GSM07,GFM09} that yields good agreement with computer simulations even for quite strong values of dissipation \cite{MSG07}. The final form \eqref{5.9} of the shear viscosity $\eta$ is obtained when one takes into account the relation \eqref{3.6}.

The kinetic parts $\kappa_k$ and $\mu_k$ of the transport coefficients characterizing the heat flux are defined, respectively, as
\begin{equation}
\label{c5}
\kappa_k=-\frac{1}{dT}\int\, d{\bf v} {\bf S}({\bf V})\cdot {\boldsymbol {\cal A}}({\bf V}),
\end{equation}
\begin{equation}
\label{c5.1}
\mu_k=-\frac{1}{dn}\int\, d{\bf v} {\bf S}({\bf V})\cdot {\boldsymbol {\cal B}}({\bf V}),
\end{equation}
where
\begin{equation}
\label{c6}
{\bf S}({\bf V})=\left(\frac{m}{2}V^2-\frac{d+2}{2}T\right){\bf V}.
\end{equation}
We obtain first the kinetic part $\kappa_k$. It is obtained by multiplying Eq.\ (\ref{5.2})
by ${\bf S}({\bf V})$ and integrating over ${\bf V}$. The result is
\begin{widetext}
\begin{equation}
\label{c7}
\left[\nu_\kappa+
\frac{1}{2}\frac{m\xi_\text{b}^2}{T}
\left(1+3\frac{\partial \zeta_0^*}{\partial \xi^*}\right)-2\zeta^{(0)}\right]\kappa_k
=-\frac{1}{dT}\int\; d{\bf V}
{\bf S}({\bf V})\cdot {\bf A},
\end{equation}
where
\begin{equation}
\label{c8}
\nu_\kappa=\frac{\int d{\bf v} {\bf S}({\bf V})\cdot {\cal L}\boldsymbol{\mathcal{A}}({\bf V})}
{\int d{\bf v}{\bf S}({\bf V})\cdot \boldsymbol{\mathcal{A}}({\bf V})}.
\end{equation}
The right hand side of Eq.\ \eqref{c7} is given by
\begin{eqnarray}
\label{c9}
-\frac{1}{d T}\int\; d{\bf V}{\bf S}({\bf V})\cdot {\bf A}&=&\frac{1}{d T}\left\{\frac{d(d+2)}{2m}nT^2\left(
1+2a_2-\frac{3}{2}\xi^*\frac{\partial a_2}{\partial \xi^*}\right)-\frac{1}{2}\int\; d{\bf V}{\bf S}({\bf V})\cdot
\boldsymbol{\mathcal{K}}\left[\frac{\partial}{\partial \mathbf{V}}\cdot \left( \mathbf{V}
f^{(0)}\right) \right]\right.\nonumber\\
& & \vicente{\left.
-\frac{3}{2}\xi^*\int\; d{\bf V}{\bf S}({\bf V})\cdot
\boldsymbol{\mathcal{K}}\left[\frac{\partial f^{(0)}}{\partial \xi^*}\right]
\right\}}\nonumber\\
&=&
\frac{1}{d T}\left\{\frac{d(d+2)}{2m}nT^2\left(
1+2a_2-\frac{3}{2}\xi^*\frac{\partial a_2}{\partial \xi^*}\right)-\frac{1}{2}\int\; d{\bf V}{\bf S}({\bf V})\cdot
\boldsymbol{\mathcal{K}}\left[\frac{\partial}{\partial \mathbf{V}}\cdot \left( \mathbf{V}
f^{(0)}\right) \right]\right.\nonumber\\
& & \vicente{\left.
-\frac{3}{2}\xi^*\frac{\partial \ln a_2}{\partial \xi^*}\int\; d{\bf V}{\bf S}({\bf V})\cdot
\boldsymbol{\mathcal{K}}\left[f^{(0)}-f_\text{M}\right]
\right\}}
\end{eqnarray}
where use has been made of Eq.\ \eqref{3.8bis} and the leading Sonine approximation \eqref{29.1} for $f^{(0)}$. The last two terms on the right hand side of Eq.\ (\ref{c9}) can be evaluated more explicitly and the result is \cite{G12}
\begin{equation}
\label{c10}
\int\; d{\bf V}{\bf S}\cdot
\boldsymbol{\mathcal{K}}\left[\frac{\partial}{\partial \mathbf{V}}\cdot \left( \mathbf{V}
f^{(0)}\right) \right]=-\frac{3}{8} 2^{d} d \frac{n T^2}{m} \chi \phi (1+\alpha)^2
\left[2\alpha-1+a_2(1+\alpha)\right],
\end{equation}
\begin{equation}
\label{c11}
\vicente{
\int\; d{\bf V}{\bf S}\cdot
\boldsymbol{\mathcal{K}}\left[f^{(0)}-f_{\text{M}}\right]=\frac{3}{32} 2^{d} d \frac{n T^2}{m} \chi \phi
(1+\alpha)^3 a_2.}
\end{equation}
\end{widetext}

With the above results, the kinetic part $\kappa_k$ can be finally written as
\begin{eqnarray}
\label{c13}
\kappa_k&=&\kappa_0\nu_0\frac{d-1}{d}\left(\nu_\kappa
+\frac{1}{2}\frac{m\xi_\text{b}^2}{T}
\left(1+3\frac{\partial \zeta_0^*}{\partial \xi^*}\right)
-2\zeta^{(0)}\right)^{-1}\nonumber\\
& &
\left\{1+2a_2-\frac{3}{2}\xi^*\frac{\partial a_2}{\partial \xi^*}+3\frac{2^{d-3}}{d+2}\phi \chi(1+\alpha)^2\right.\nonumber\\
&\times &\vicente{  \left.\left[2\alpha-1+a_2(1+\alpha)
-\frac{3}{8}(1+\alpha)\xi^*\frac{\partial a_2}{\partial \xi^*}\right]\right\}},
\end{eqnarray}
where $\kappa_0$ is the low density value of the thermal conductivity of an elastic gas (defined by Eq.\ \eqref{5.13}).

\vicente{In order to get an explicit expression for $\kappa_k$}, one considers the form \eqref{31} for $\zeta^{(0)}$ and evaluates $\nu_\kappa$ by considering again the leading terms in a Sonine polynomial expansion of $\boldsymbol{\mathcal{A}}({\bf V})$. With these approaches, one gets the expression \eqref{5.15} for $\nu_\kappa$ while
\begin{equation}
\label{c14.1}
\frac{\partial \zeta_0^*}{\partial \xi^*}=\zeta_\text{M} \frac{\partial a_2}{\partial \xi^*},
\end{equation}
where $\zeta_\text{M}$ is defined by Eq.\ \eqref{5.14.1}. Use of \eqref{c14.1} in Eq.\ \eqref{c13} gives the final result.

In order to determine $\mu_k$, we multiply Eq.\ \eqref{5.3} by ${\bf S}({\bf V})$ and integrate over velocity to get
\begin{widetext}
\begin{eqnarray}
\label{c14.2}
\left[\nu_\mu-\frac{3}{2}\left(\zeta_\text{s}^{(0)}
-\frac{m\xi_\text{b}^2}{T_\text{s}}\right)\right]\mu_k&=&-\frac{1}{d n}\int\; d{\bf V}
{\bf S}({\bf V})\cdot \left\{\zeta ^{(0)}\left(1+\phi\partial_{\phi}\ln \chi \right)\boldsymbol{\mathcal{A}}
\right.\nonumber\\
& & \left.+\left[\phi \frac{\partial \chi}{\partial \phi}\frac{\partial}{\partial \chi}\left(\frac{\zeta^{(0)}}{\chi}\right)-\xi^*\frac{\partial \zeta^{(0)}}{\partial \xi^*}
-\frac{2}{3}\theta\frac{\partial \zeta^{(0)}}{\partial \theta}\right]\boldsymbol{\mathcal{A}}
+{\bf B}\right\}\nonumber\\
&=&\frac{T}{n}\left[\zeta ^{(0)}\left(1+\phi\partial_{\phi}\ln \chi \right)+\phi \frac{\partial \chi}{\partial \phi}\frac{\partial}{\partial \chi}\left(\frac{\zeta^{(0)}}{\chi}\right)-\xi^*\frac{\partial \zeta^{(0)}}{\partial \xi^*}
-\frac{2}{3}\theta\frac{\partial \zeta^{(0)}}{\partial \theta}\right]\kappa_k\nonumber\\
& & 
-\frac{1}{d n}\int\; d{\bf V}{\bf S}({\bf V})\cdot {\bf B},
\end{eqnarray}
where
\begin{equation}
\label{c14.2bis}
\nu_\mu=\frac{\int d{\bf v} {\bf S}({\bf V})\cdot {\cal L}\boldsymbol{\mathcal{B}}({\bf V})}
{\int d{\bf v}{\bf S}({\bf V})\cdot \boldsymbol{\mathcal{B}}({\bf V})}.
\end{equation}
\vicente{
The last term on the right hand side of Eq.\ \eqref{c14.2} is
\begin{eqnarray}
\label{c14.3} -\frac{1}{dn}\int\;d{\bf V}{\bf S}\cdot {\bf B}&=&\frac{d+2}{2}\frac{T^2}{m}\left(a_2
-\xi^*\frac{\partial a_2}{\partial \xi^*}
-\frac{2}{3}\theta\frac{\partial a_2}{\partial \theta}+\phi \frac{\partial \chi}{\partial \phi}
\frac{\partial a_2}{\partial \chi}\right)\nonumber\\
& & -\frac{3}{32} 2^{d}\frac{T^2}{m}\chi \phi (1+\alpha)^3\left(\xi^*\frac{\partial a_2}{\partial \xi^*}
+\frac{2}{3}\theta\frac{\partial a_2}{\partial \theta}-\phi \frac{\partial \chi}{\partial \phi}
\frac{\partial a_2}{\partial \chi}\right)\nonumber\\
& & +\frac{1}{dn}
\left(1+\frac{1}{2}\phi\partial_\phi\ln \chi\right)\int\;d{\bf V}{\bf S} \cdot
\boldsymbol{\mathcal{K}}\left[f^{(0)}\right].
\end{eqnarray}
}
\end{widetext}
The last term in Eq.\ \eqref{c14.3} is given by
\begin{eqnarray}
\label{c14.4}
\int\; d{\bf V}{\bf S}\cdot
\boldsymbol{\mathcal{K}}\left[f^{(0)}\right]&=&\frac{3}{8} 2^{d} d \frac{n T^2}{m} \chi \phi (1+\alpha)
\left[\alpha(\alpha-1)\right.\nonumber\\
& & \left.+\frac{a_2}{6}\left(10+2d-3\alpha+3\alpha^2\right)\right],\nonumber\\
\end{eqnarray}
The final expression of $\mu_k$ is obtained from Eq.\ \eqref{c14.2} when one substitutes Eq.\ \eqref{c14.3} into Eq.\ \eqref{c14.2}. However, this expression is not explicit unless one knows the collision frequency $\nu_\mu$. To determine it, one takes the leading terms in a Sonine polynomial expansion of $\boldsymbol{\mathcal{B}}({\bf V})$ and gets $\nu_\mu=\nu_\kappa$. This finally yields Eq.\ \eqref{5.17} for $\mu_k$.

We consider finally the first-order contribution $\zeta_U$ to the cooling rate. This coefficient is given by Eq.\ \eqref{5.19}, where
\begin{eqnarray}
\zeta_{11}&=&\frac{1}{2nT}\frac{\pi ^{(d-1)/2}}{d\Gamma \left( \frac{d+3}{2} \right)}
\sigma ^{d-1}\chi m (1-\alpha^{2})\nonumber\\
&\times & \int d\mathbf{V}_{1}\,\int d
\mathbf{V}_{2}\,g^{3}f^{(0)}(\mathbf{V}_{1})\mathcal{D}(\mathbf{V}_{2}), \label{c15}
\end{eqnarray}
and the unknown $\mathcal{D}$ verifies the integral equation \eqref{5.5}. An approximate solution to \eqref{5.5} can be obtained by taking the Sonine approximation
\begin{equation}
\mathcal{D}(\mathbf{V})\rightarrow e_{D}f_{M}(\mathbf{V})F( \mathbf{V}), \label{c16}
\end{equation}
where
\begin{equation}
F(\mathbf{V})=\left( \frac{m}{2T}\right) ^{2}V^{4}-\frac{d+2}{2}
\frac{m}{T}V^{2}+\frac{d(d+2)}{4}, \label{c17}
\end{equation}
and the coefficient $e_D$ is
\begin{equation}
e_{D}=\frac{2}{d(d+2)}\frac{1}{n}\int \;d\mathbf{V}\;\mathcal{D}(
\mathbf{V})F(\mathbf{V}). \label{c18}
\end{equation}
Substitution of Eq.\ (\ref{c16}) into Eq.\ (\ref{c15}) gives
\begin{equation}
\label{c19}
\zeta_{11}=\frac{3(d+2)}{32d}\chi (1-\alpha^2)
\left(1+\frac{3}{32}a_2\right)\nu_0 e_D.
\end{equation}
The coefficient $e_D$ is determined by substituting Eq.\ (\ref{c16}) into the integral
equation (\ref{5.5}), multiplying by $F({\bf V})$ and integrating over ${\bf V}$. After some algebra one gets the
expression \eqref{5.21} for $\zeta_{11}$. Here, for the sake of simplicity, we have neglected the contributions proportional to
the derivatives $\partial a_2/\partial \theta$ and  $\partial a_2/\partial \chi$.


\section{Expressions for choice B}
\label{appD}

In this Appendix we display the expressions for the Navier-Stokes transport coefficients $\eta$, $\lambda$, $\kappa$, and $\mu$ by using the choice B defined by the condition $\partial_t^{(0)}T=0$. The application of the Chapman-Enskog method to this case follows similar mathematical steps as those made for the choice A ($\partial_t^{(0)}T\neq 0$). The results show that the expressions of $\eta$ and $\lambda$ are the same as those obtained for the choice A
[see Eqs.\ \eqref{5.7}--\eqref{5.9}]. However, the forms of $\kappa$ and $\mu$ are different since they
are given by Eqs.\ \eqref{5.12} and \eqref{5.16}, respectively, but their corresponding kinetic contributions are
\begin{eqnarray}
\label{5.24}
\kappa_k&=&\frac{d-1}{d}\frac{\kappa_0\nu_0}{\nu_\kappa+\frac{3\beta m}{T_\text{b}}\xi_\text{b}^2}
\left(1+2a_{2,\text{s}}\right.\nonumber\\
&+& \left.3\frac{2^{d-3}}{d+2}\phi \chi(1+\alpha)^2\left[2\alpha-1+a_{2,\text{s}}(1+\alpha)\right]\right),\nonumber\\
\end{eqnarray}
\begin{eqnarray}
\label{5.25}
\mu_k&=&\frac{\kappa_0\nu_0T_\text{s}}{\nu_\kappa+\frac{3\beta m}{T_\text{b}}\xi_\text{b}^2}\left\{\frac{d-1}{d}a_2
+3\frac{2^{d-2}(d-1)}{d(d+2)}\phi \chi
(1+\alpha)\right.\nonumber\\
&\times&
\left(1+\frac{1}{2}\phi\partial_\phi\ln
\chi\right)\left[\alpha(\alpha-1)\right.\nonumber\\
& & \left.\left.+
\frac{a_{2,\text{s}}}{6}(10+2d-3\alpha+3\alpha^2)\right]\right\},
\end{eqnarray}
where the collision frequency $\nu_\kappa$ is defined by Eq.\ \eqref{5.15}.

\end{document}